\documentclass[pdflatex,sn-mathphys-num]{sn-jnl}

\usepackage{graphicx}%
\usepackage{multirow}%
\usepackage{amsmath,amssymb,amsfonts}%
\usepackage{amsthm}%
\usepackage{mathrsfs}%
\usepackage[title]{appendix}%
\usepackage{xcolor}%
\usepackage{textcomp}%
\usepackage{manyfoot}%
\usepackage{booktabs}%
\usepackage{algorithm}%
\usepackage{algorithmicx}%
\usepackage{algpseudocode}%
\usepackage{listings}%
\usepackage{moreverb,url}
\usepackage{hyperref}
\usepackage[capitalise,noabbrev]{cleveref}
\usepackage{caption,subcaption}

\usepackage[export]{adjustbox}
\usepackage{tabularx}
\usepackage{bm}
\usepackage{soul}
\usepackage{multicol}
\usepackage{adjustbox}
\usepackage{array}
\usepackage{pdfpages}

\theoremstyle{thmstyleone}%
\theoremstyle{thmstyletwo}%

\theoremstyle{thmstylethree}%

\begin{document}
\includepdf[pages=1]{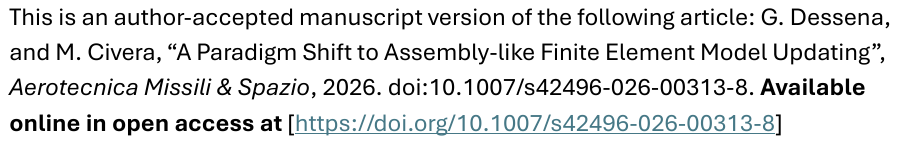}
\title[A paradigm shift to assembly-like finite element model updating]{A paradigm shift to assembly-like finite element model updating\footnote{\hl{This is an author-accepted manuscript version of the following article: G. Dessena, and M. Civera, “A Paradigm Shift to Assembly-like Finite Element Model Updating”, \textit{Aerotecnica Missili \& Spazio}, 2026. doi:10.1007/s42496-026-00313-8. \textbf{\textit{Available online in open access at}}} [\url{https://doi.org/10.1007/s42496-026-00313-8}]}}


\author*[1,2]{\fnm{Gabriele} \sur{Dessena}}\email{GDessena@ing.uc3m}

\author[3]{\fnm{Alessandro} \sur{Pontillo}}\email{Alessandro.Pontillo@uwe.ac.uk}

\author[1]{\fnm{Dmitry I.} \sur{Ignatyev}}\email{D.Ignatyev@cranfield.ac.uk}
\author*[1]{\fnm{James F.} \sur{Whidborne}}\email{J.F.Whidborne@cranfield.ac.uk}
\author[1]{\fnm{Luca} \sur{Zanotti Fragonara}}\email{L.ZanottiFragonara@cranfield.ac.uk}

\affil[1]{\orgdiv{School of Aerospace, Transport and Manufacturing}, \orgname{Cranfield University}, \orgaddress{\street{College Road}, \city{Cranfield}, \postcode{MK430AL}, \state{Bedfordshire}, \country{UK}}}

\affil[2]{\orgdiv{Department of Aerospace Engineering}, \orgname{Universidad Carlos III de Madrid}, \orgaddress{\street{Avenida de la Universidad 30}, \city{Leganés}, \postcode{28911}, \state{Madrid}, \country{Spain}}}

\affil[3]{\orgdiv{School of Engineering}, \orgname{University of the West of England}, \orgaddress{\street{Frenchay Campus, Coldharbour Lane}, \city{Bristol}, \postcode{BS16 1QY}, \country{UK}}}


\abstract{In general, there is a mismatch between a finite element model {(FEM)} of a structure and its real behaviour. In aeronautics, this mismatch must be small because {FEM}s are a fundamental part of the development of an aircraft and of increasing importance with the trend to more flexible wings in modern designs. {Iterative f}inite element model updating can be computationally expensive for complex structures, and surrogate models can be employed to reduce the computational burden. A novel approach for {FEM} updating, namely assembly-like, is proposed and validated using real experimental data {from a flexible wing.} The assembly-like model updating framework implies that the model is updated as parts are assembled. {Benchmarking against the classical global, or one-shot, approach demonstrates that the proposed method is more computationally efficient, since a normalised workload proxy based on solver-reported model size and memory footprint indicates about 28\% lower overall effort. Aapproximately 95\% of the required solves are performed on lower-fidelity subassembly models with smaller equation counts and memory requirements. Despite the reduced reliance on full-wing evaluations, the new approach retains the fidelity, within 1\% of a joint natural frequencies and modal shapes index, of the global approach.}}




\maketitle
\section*{Nomenclature}
\subsection*{Definitions, Acronyms and Abbreviations}
\begin{multicols}{2}
\noindent A: Area \\ 
AR: Aspect Ratio \\ 
APDL: Ansys Parametric Design Language \\
BearDS: Beam Reduction Dynamic Scaling \\ 
$E$: Young Modulus \\ 
E$_exp$: Experimental Data \\
EGO: Efficient Global Optimization \\ 
$\bm{C}$: Damping Matrix \\ 
$\bar{c}$: Mean Aerodynamic Chord \\ 
EGO: Efficient Global Optimization \\ 
EI: Expected Improvement\\ 
$EI$: Bending Stiffness \\ 
EFWE: Equivalent Full-Wing Evaluations \\
EMA: Experimental Modal Analysis \\ 
FE: Finite element
FEM: Finite Element Model \\ 
FEMU: Finite Element Model Updating \\ 
FEMU\_1: FEM Updated via Bottom-up \\ 
FEMU\_2: FEM Updated via Top-down \\ 
GA: Genetic Algorithm \\ 
GVT: Ground Vibration Testing \\ 
HAR: High Aspect Ratio \\ 
$I$: Moment of Inertia \\ 
J: Torsion Constant \\ 
L: Length \\
$m$: One of the FEM models: spar, torque box, and full-wing (XB-2)
$\bm{M}$: Mass Matrix \\ 
$M_m$: solver-reported total allocated memory \\
MAC: Modal Assurance Criterion \\ 
MTMAC: Modified Total Modal Assurance Criterion \\ 
$n$: Mode number \\
$N_\mathrm{eq,m}$: number of FE solver calls for model $m$
$N_\mathrm{eval,m}$: number of FE solver calls for model $m$ \\
N$_num$: Numerical data \\
P: Static Load \\
rEGO: refined Efficient Global Optimisation \\ 
RMS: Root Mean Square \\ 
XB-2: eXperimental BeaRDS-2 \\ 
$x_n$: Optimisation Variable \\ 
$\alpha$: Mass Proportional Rayleigh Damping Coefficient \\ 
$\beta$: Stiffness Proportional Rayleigh Damping Coefficient \\ 
$\delta_{tip}$: Tip {Deflection}\\
$\zeta_n$: Damping Ratio \\ 
$\lambda$: Taper Ratio \\ 
$\Lambda_{c/4}$: Quarter-chord Sweep Angle\\ 
$\Lambda_\text{LE}$: Leading Edge Sweep Angle\\ 
$\rho$: Density \\ 
$\bm{\phi}_n$: Mode Shape \\ 
$\omega_n$: Natural Frequency \\ 
\end{multicols}

\section{Introduction}
Over the last three decades, increasing computational power has allowed for the rapid development of finite element model updating (FEMU) methods \cite{Cao2023}. This is because a general mismatch between the finite element models (FEMs) and the real systems they are meant to describe usually exists \cite{Tao2025}.

In \cite{Alkayem2018}, FEMU techniques are divided into two categories: direct and indirect methods. The former is not suitable for practical engineering applications as (i) they require very precise measurements of the structural vibration response, (ii) have a high sensitivity to noise, (iii) cannot be used with truncated data and (iv) are prone to lose symmetry in the FEM matrix. {Nevertheless, some practical applications exist in the literature. Direct methods include the direct matrix approach, which performs a single-step update via matrix mixing (a modal outer-product expansion of the inverse of the mass and stiffness matrices), as shown in }\cite{Jull2019}{ for a steel T-structure model. In contrast, eigenstructure assignment methods update existing system matrices through state-feedback control that reassigns eigenvalue–eigenvector pairs using a gain matrix, thereby modifying the transient response. This is demonstrated in }\cite{Sen2013}{ on two numerical FEM.} However, indirect, or iterative, methods accommodate these drawbacks. Moreover, iterative methods, driven by the minimisation of penalty functions, can require a heavy computational burden. This happens for evolutionary techniques, such as genetic algorithms (GAs) \cite{Dessena2022d}. Hence, surrogate-based techniques can be employed for more efficient use of computational power. This is done in \cite{Yang2017} for the FEMU of numerical and benchmark structures using the well-known efficient global optimization (EGO) on frequency domain data. In order to improve the local capabilities of a global algorithm like EGO, an enhanced version of EGO, the refined efficient global optimisation (rEGO) has been proposed in \cite{Dessena2022c,Dessena2022d}. rEGO was successfully applied to numerical and experimental systems for damage detection via FEMU. 
The reader interested in a more comprehensive review on FEMU can refer to \cite{Mottershead1993} and~\cite{Friswell1995}, while for thorough reviews of {indirect} methods, the works of \cite{Marwala2010} and \cite{Alkayem2018} are suggested.

The FEMU task is important across all fields of engineering, finding applications such as damage detection \cite{Shadan2018}, identification \cite{Forouzesh2023}, topology optimisation \cite{Lu2025}, and virtual testing \cite{Peri2025}. FEMU is also crucial in aeronautics, where having a reliable FEM is pivotal in meeting certification requirements \cite{DeFlorio2011}. {Further} challenges {have} recently {arisen} for FEMU in aeronautics, in particular for aircraft wings, {because} the design paradigm is shifting towards lightweight materials and slenderer wings, {in order to improve} their aerodynamic efficiency \cite{Malik2020}. Hence, developing accurate models is not only necessary for {analysing} aeroelastic effects, but also for design and controls. In particular, \cite{DiLeone2021} develops a FEM of a scaled aircraft model to be updated with {ground vibration testing (GVT)} data from different loaded scenarios to estimate flutter onset speed. In \cite{Mihaila-Andres2019a} a FEMU technique is implemented for positioning {a} wing box composite material layers for passive aeroelastic suppression, while in \cite{Wang2018} model updating is used for enhancing the handling qualities of an aircraft with flexible wings. FEMU has also been implemented for other applications, such as reverse engineering of a fighter aircraft internal structure \cite{Chiodi2021}. Nevertheless, FEMU, in aeronautics, is not only carried out on wings, but also on the full aircraft, {e.g.} \cite{Cecrdle2022}, and components, {e.g.} \cite{Wang2020}.

Experimental data is required to update a FEM. In most cases, results from vibration data are used \cite{Das2024}. The most prominent product of vibration data are modal parameters \cite{Dessena2022}, which can be obtained {by experiment}, the so-called experimental modal analysis (EMA) \cite{Dessena2022f}, and during normal operation, {known as} operational modal analysis \cite{Lu2022}. A particular type of EMA, used in aeronautics, is GVT and in this work, the two terms are used interchangeably as {the work} deals with aeronautical structures. In fact, modal parameters, in particular, natural frequencies ($\omega_n$) and mode shapes ($\bm{\phi}_n$) are a common metric for FEMU \cite{Jalali2022}. Notably, \cite{Coppotelli2019} used $\bm{\phi}_n$, via the modal assurance criterion (MAC) \cite{Allemang1982}, from operational data for the FEMU of a wing section.

 Having an accurate structural model of an aircraft structure is not fundamental only for the structural assessment per se, but also for the general life-cycle assessment of the aircraft. Recent advances in the digital twin paradigm have shown that this is vital for building reliable models \cite{Tavares2024}. These challenges become tougher nowadays as wings become more flexible. For example, \cite{Sharqi2022,Sharqi2023} propose a method {for the numerical extraction of modal data} from deformed wings (vertical displacement of the wing); however, this was not further explored experimentally. In \cite{Zhao2017} a component-based FEMU strategy is applied to a composite wing aircraft. Instead of updating the {complete} structure, {the updating is carried out for an individual component, before assembly into the whole system; however, no comparison is given to one-shot approaches, neither in terms of performance nor precision.} The latter is addressed, on the same system, in \cite{Zhao2019}, showing favourable results, computationally speaking, for the progressive approach. However, the implementation suggested by Zhao in the aforementioned works{, in particular their intermediate steps,} is more similar to the {hybrid testing approach} \cite{McCrum2016a,Yang2020}, as usually defined within Civil Engineering, where only part of a structure, or system {(e.g. the port wing)}, is known{, rather than to an assembly-like procedure. In fact, EMA data is available only for the wing and for the full assembly, rather than for all the components.} Notably, a similar application {is carried out for the structural health monitoring of a suspended bridge model} in \cite{Wilson2023}. {Moreover, a substructure-based {FEMU} approach is introduced in }\cite{Weng2011}{, wherein selected eigenmodes and derivatives from independent substructures are assembled into a reduced eigenequation, allowing global eigensolutions to be recovered with re-analysis restricted to the relevant substructures.} Nevertheless, from these works it became apparent that the possibility of carrying out assembly-like model assembly, {and} updating, is appealing as one can consider smaller, in terms of computational loads (e.g. nodes and elements), models, rather than larger and more complex ones. However, a more dedicated approach to the paradigm shift from one-shot and partial model approaches to progressive assembly-like approaches is needed{, since these works have not addressed the stepwise characterisation of all subassemblies.}

In order to establish the feasibility, goodness and efficiency of this method, a comparison between the proposed {assembly-like} and the traditional direct approaches is needed. {This comparison verifies performance in terms of model improvement and efficiency in terms of (i) a normalised, non-time-based cost proxy, the Equivalent Full-Wing Evaluations (EFWE), which aggregates model-evaluation counts with solver-reported model complexity (active DOFs/equations) and allocated memory, and (ii) the distribution of model evaluations across subassembly versus full-assembly models required for convergence.}

{Given the aforementioned motivation, the novelty of this work is to propose and assess a framework for the assembly-like FEMU against the classical direct approach to show its goodness and efficiency.} 

{The practical impact of the proposed framework is to enable more accurate and computationally efficient FEMU for assemblable structures, provided that experimental data can be obtained at the subassembly (and, where feasible, component) level. The main practical requirement is therefore testability across assembly stages, including the availability of appropriate excitation/measurement access and the possibility of reconfiguring the experimental setup to characterise subassemblies individually. In addition, the current study assumes predominantly linear structural dynamics, such that the identified modal information can be compared consistently with the linear FEM and used within the adopted updating objective. From an application perspective, the approach is most attractive when the added experimental effort required to test subassemblies does not offset the computational savings achieved by reducing the number of evaluations of the most expensive full-assembly model. As a result, near-term use cases are expected to include engineering systems where subassemblies are accessible and readily testable, such as aeroelastic demonstrators (e.g., flexible wings) and small unmanned aircraft (up to 25~kg maximum take-off mass).}

{The proposed framework differs from the component-assisted, multi-level optimisation strategy in }\cite{Zhao2019} {in three main respects. 
First, the present work follows a fully progressive, assembly-level workflow, in which each subassembly is updated using its own experimental dataset and the resulting updated parameters are propagated to the next (larger) assembly stage; this is enabled here by the availability of experimental data for the parts, subassemblies, and the full assembly. 
Second, the updating at each stage is formulated on a consistent set of physical parameters that affect mass and stiffness together for the considered subassembly, rather than splitting the problem into separate mass-then-stiffness optimisation levels. 
Third, the optimisation is carried out using a surrogate-based strategy (rEGO) that treats the finite element (FE) solver as a black box and is designed to reduce the number of evaluations of the most expensive model, which is different in scope from the deterministic multi-level optimisation adopted in }\cite{Zhao2019}. 
{The proposed framework also differs from substructure-based approaches such as }\cite{Weng2011}{. There, the structure is partitioned to accelerate the computation of global eigensolutions and eigensensitivities by assembling reduced eigenequations and re-analysing only the substructures that are relevant to the current optimisation step. In contrast, the present work is not a reduced-eigenproblem or eigensensitivity acceleration technique; instead, it is an assembly-driven FEMU workflow that updates physically testable subassemblies in a stepwise manner and then updates the final assembly accordingly.}

The framework is tested on a flexible wing model, of which thorough experimental data, involving the full wing and its parts and sub-assemblies, is available \cite{Dessena2022b,Dessena2022a}. The hypothesis is that there could be an improvement in model accuracy and computational performance compared to {traditional} one-shot approaches. {The model updating procedure followed in this work is based on a surrogate-based iterative method based on rEGO, which aims at minimising the Modified Total Modal Assurance Criterion (MTMAC)} \cite{Perera2006} {to match the discretised FEM modal response to experimental results. {MTMAC combines natural frequency consistency and mode-shape correlation in a multiplicative manner, such that inadequate agreement in any selected mode is not masked by good agreement in the remaining modes. Additionally, this avoids introducing ad-hoc weighting factors between frequency- and mode shape-based residuals.} Nevertheless, any other iterative optimiser and/or valid FEMU goal function could be implemented with the same concept.}

The {aim} of the work is to prove that the proposed, assembly-like FEMU, novel method is more computationally efficient when compared to the classical approach. Thus, the {objectives} of this work are the following:
\begin{enumerate}
    \item[(i)] {Propose and assess the merits of the assembly-like approach;}
    \item[(ii)] {Obtain an updated discretised FEM of a complex model with the approach in (i).}
\end{enumerate}

{The remaining sections of this article are organised as follows:}
\begin{enumerate}
    \item[2.] \hyperref[sec:met]{\emph{Materials and Methods}}{: The flexible wing model is introduced, the experimental campaign described, the FEMs discretisation discussed, and the FEMU workflow is outlined;}
    \item[3.] \hyperref[sec:res]{\emph{Results}}{: The results from the proposed FEMU method are presented;}
    \item[4.] \hyperref[sec:dis]{\emph{Discussion}}{: The computational efficiency, overall precision, and limitations of the assembly-like FEMU are discussed;}
    \item[5.] \hyperref[sec:7con]{\emph{Conclusions}}{: Final remarks and a summary of the results are presented.}
\end{enumerate}

\section{Materials and Methods}\label{sec:met}


\subsection{The Flexible Wing Model\label{sec:7xb2}}
\begin{figure*}[!ht]
\centering
    \begin{subfigure}[t]{.8\textwidth}
	\centering
		\includegraphics[width=.89\textwidth,keepaspectratio]{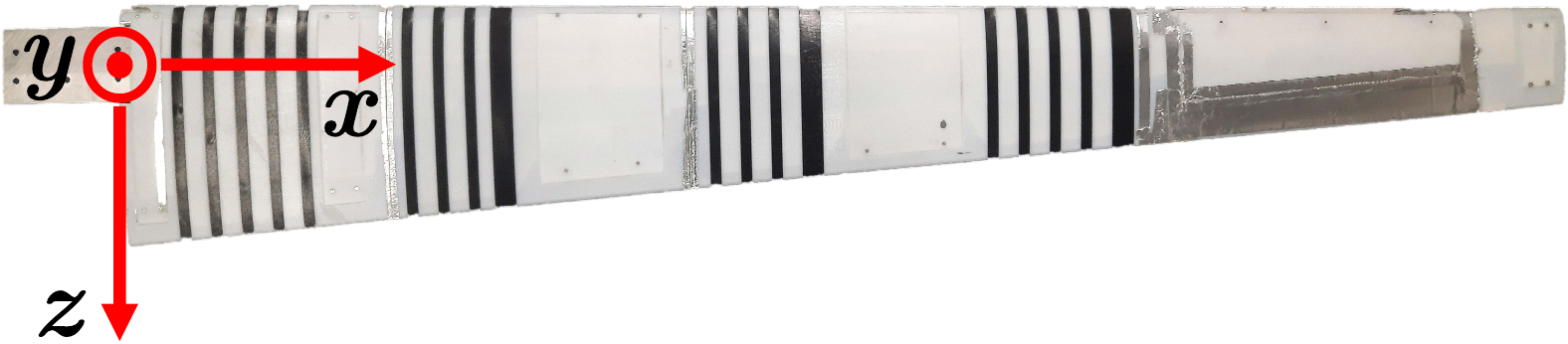}
		\captionsetup{font={it},justification=centering}
		\subcaption{\label{fig:7xb2c}}	
	\end{subfigure}
    \begin{subfigure}[t]{.8\textwidth}
	\centering
		{\includegraphics[width=\textwidth,keepaspectratio]{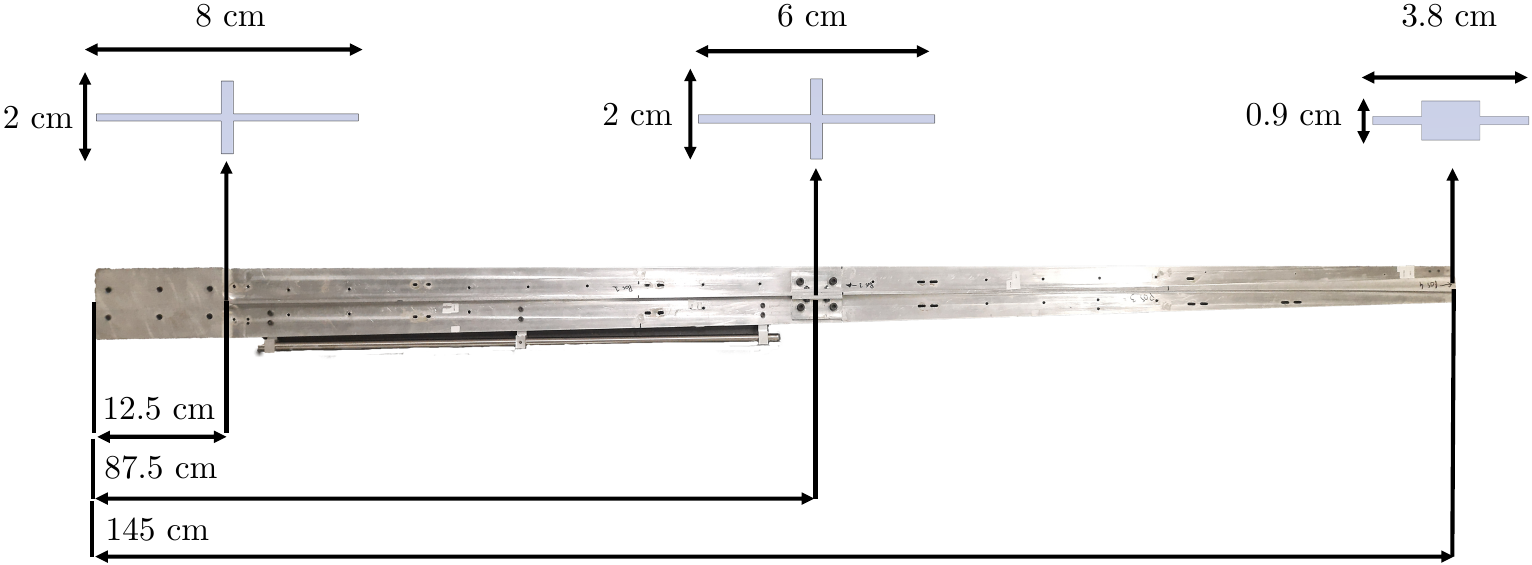}}
		\captionsetup{font={it},justification=centering}
		\subcaption{\label{fig:7xb2b}}	
	\end{subfigure}
	\caption{XB-2: Top views of the full wing (\cref{fig:7xb2c} - axes reference system for the $\bm{\phi}_n$ plot (Adapted from \cite{Dessena2022g})) and the torque box (\cref{fig:7xb2b} - Adapted from \cite{Dessena2023}). \Cref{fig:7xb2b} also features the spar cross-sections and information about their span-wise position. Not to scale.}
	\label{fig:7xb2}
\end{figure*}

The wing model chosen to validate the FEMU technique is a high aspect ratio (HAR) flexible wing model developed at Cranfield University for the Beam Reduction Dynamic Scaling (BeaRDS) project \cite{Pontillo2018}. The main aim of BeaRDS was to establish a workflow for the design, manufacture and testing of dynamically scaled HAR wings for use in Cranfield University 8$\times$6 {ft (2.4$\times$1.8 m)} wind tunnel.
{Specifically, the wing under scrutiny is the eXperimental BeaRDS-2 (XB-2) model (}\cref{fig:7xb2c}{). The wing model has been dynamically scaled to an aircraft similar to the Airbus A320, maintaining the same MTOW but with a slenderer wing (higher AR), the same wing load, and a cruise speed of M = 0.6 to represent an efficient turboprop option. The wing span, optimised using eXergy methods }\cite{Hayes2019}{, measured 48 m. To fit in the wind tunnel, a scaling factor of 16 was established. The full-scale model was non-dimensionalised by matching the aeroelastic equations for non-dimensional mass and frequencies. More details on this process are available in} \cite{Yusuf2019,Pontillo2020}.
The scaled model is made up of four main components: the spar, the stiffening tube, the additional brass ballasts and the skin. The spar and stiffening tube serve as the wing torque box (\cref{fig:7xb2b}), the brass ballasts were added for dynamic scaling purposes \cite{Yusuf2019t}, and the skin was designed to minimise its effect on the wing stiffness \cite{Yusuf2019}. Since dynamic scaling is not the focus of this work, the brass ballasts are removed for the current study.
The spar is machined from two 6082-T6 Aluminium Alloy blocks and joined at mid-span with a weld and reinforced by four bolted L-section plates. Geometrically the spar can be divided into three main sections, according to its profile, as shown in \cref{fig:7xb2c}. The spar mass is 1.225 kg.

A stiffening tube was added aft of the spar to postpone flutter onset during the wind tunnel test of the BeaRDS project \cite{Yusuf2019t}. The tube is linked to the main spar at three points (near {each} end and in the middle), as shown in \cref{fig:7xb2c}.
The tube is made of stainless steel and features an outer diameter of 10 mm, with a thickness of 1 mm and a length of 600 mm. The torque box, consisting of the spar and the tube assembly, has a combined mass of 1.362 kg.

The outer surface of the wing is defined by the skin (\cref{fig:7xb2c}), made up of 47 different subsections, which are 3D printed in Digital ABS (white) and Agilus 30 (black), a rubber-like material. Despite the use of different materials for the adjacent strips, the skin is made only of three separate parts, thanks to the PolyJet technology, which allows for the 3D printing of subsequent layers with different materials~\cite{Macdonald2017}. This, and the use of a rubber-like material, gives enhanced flexibility to the wing skin, allowing for large tip displacements. The assembly of the torque box and the skin constitutes the full wing. \Cref{tab:7prop} introduces the wing material and physical properties, including aerofoil, aspect ratio (AR), mean aerodynamic chord ($\bar{c}$), taper ratio ($\lambda$), Leading Edge sweep ($\Lambda_\text{LE}$) and quarter-chord sweep ($\Lambda_{c/4}$). Please note, the wing has neutral twist and dihedral angles.

\begin{table}[!ht] 
\caption{{Material, physical, and geometric properties.}\label{tab:7prop}}
\centering
\begin{tabular}{llll}
\toprule
\textbf{Material}	& {\textbf{E }- Young Modulus [GPa]}	& {\textbf{Poisson Ratio }[-]} & $\bm{\rho}$ - Density [kgm]$^{-3}$]\\\midrule
 6082-T6 Aluminium		& 70			& 0.33 & 2700\\
Stainless Steel		& 193			& 0.33 & 8000\\
Digital ABS		& 2.6--3.0			& 0.33 \cite{keane2017} & 1170--1180\\
Agilus 30		& 3e-3 \cite{Dykstra2022}			& 0.40 \cite{Bossart2021} & 1140\\\midrule
\textbf{Property}	& \textbf{Details}	& {\textbf{Unit}}&\\\midrule
Semi span & 1.5 & m &\\
AR & 18.8 & - \\
$\bar{c}$ & {0.172} & {m} & \\
$\lambda$ & 0.35 & - & \\
$\Lambda_\text{LE}$ & 14.9 & $^\circ$ &\\
$\Lambda_{c/4}$ & 0 & $^\circ$ & \\
Aerofoil & NACA 23015 & - & \\
Mass & 3.024 & kg &\\\botrule
\end{tabular}
\end{table}

The reader interested a more profound review of the BeaRDS project is referred to \cite{Pontillo2018,Yusuf2019,Hayes2019,Pontillo2020,Yusuf2019t} and further information on the XB-2 wing can be found in \cite{Pontillo2020,Dessena2022b}.

\subsubsection{Preliminary Finite Element Models}
Having outlined the general geometry, components and properties of the XB-2, three preliminary discretised FEMs are built in ANSYS Mechanical APDL 2021\footnote{{APDL stands for ANSYS Parametric Design Language. See here for more information on ANSYS Mechanical APDL  2021:}\url{https://storage.ansys.com/mbu-assets/Mech/whatsnew/v212/rev4/WhatsNew.htm}} {(the choice of using the ANSYS APDL environment is clarified later, in }\cref{sec:7femu}). Respectively, a spar, torque box and full wing are built following an {{assembly-like}} approach. First, the spar model is built, then the tube is added for the torque box model and finally, the wing skin is modelled. This approach {can be} followed because experimental data of each scenario is available in \cite{Dessena2022b}.

\begin{figure*}[!ht]
\centering
		{\includegraphics[width=.8\textwidth,keepaspectratio]{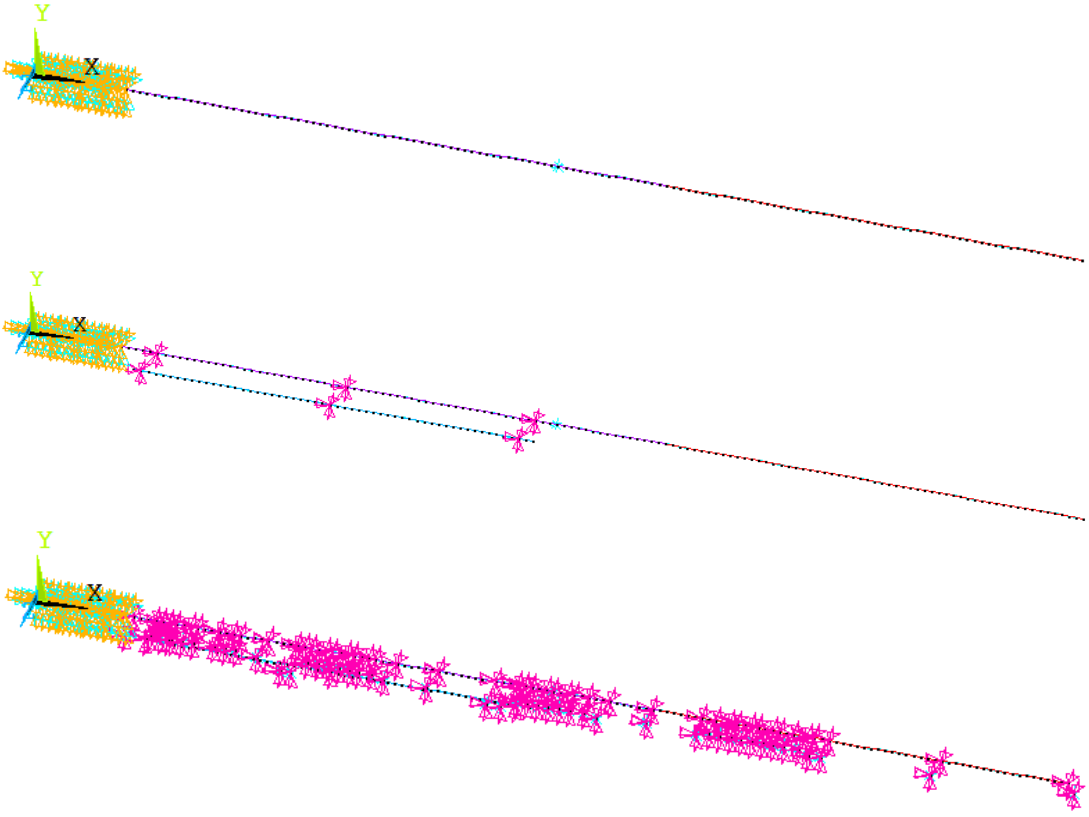}}
	\caption{FEMs of the spar (top), torque box (middle) and full wing (bottom).}
	\label{fig:fem}
\end{figure*}

The spar is easily discretised as a beam with 3 different {portions}. The first section is rectangular and represents the clamped root. Then three sections are defined along the spar to represent the three different {section profile changes; hence, defining the two tapered sections along the span.} The reinforcement plates are modelled as a 63 g lumped mass at mid-span. \verb|BEAM188| and \verb|MASS21| elements are used to, respectively, model the spar and the reinforcement plates. 6082-T6 Aluminium is assigned to the three sections. The \verb|BEAM188| element is based on Timoshenko beam theory with a first-order shear-deformation theory, while \verb|MASS21| element represents a single-node concentrated mass with components in the element coordinate directions. In \cref{fig:fem} (top){,} the FEM of the spar is shown with its boundary conditions. {The end of the spar is fully, all DOFs, constrained, in the cantilever sense,} and is identified by the yellow arrows{, while {the reinforcement} plates,} discretised as a lumped mass (\verb|MASS21|), are shown as a cyan asterisk.

Building on the spar model, the torque box FEM is constructed by adding the stiffening tube aft of the existing spar. The tube is discretised with a \verb|BEAM188| element and it is linked to the spar with 3 rigid link constraints (\verb|CERIG| - rigid constraints between a master and slave node) for all degrees of freedom, shown in magenta in \cref{fig:fem} (middle). Stainless steel material properties are assigned to the tube. 

Finally, the full wing model is created by adding the skin, as shown in \cref{fig:fem} (bottom). The skin is designed \emph{to limit the impact of the skin on the model overall stiffness} \cite{Pontillo2020}. This statement is verified numerically with solid-element FEMs of the skin and the torque box assembly in ANSYS Workbench Mechanical. The two FEMs are used to obtain the tip deflection ($\delta_{tip}$) from a static force input ($P$) at the {tip end face} of 1 N. Then, by assuming the standard relationship from Euler–Bernoulli analysis of uniform cantilever beams:
\begin{equation}
\delta_{tip}=\frac{PL^3}{3EI}\ =>EI=\frac{PL^3}{3\delta_{tip}}    
\end{equation}
{t}he results in \cref{tab:r1_stiff} are obtained, which prove that the bending stiffness, $EI$, of the skin is negligible if compared to that of the torque box, as the latter is only 1\% of the former.
\begin{table}[!ht] 
\caption{Results of the static analyses in ANSYS Workbench Mechanical for the torque box and skin. Only the suspended lengths are considered, without the clamped root.\label{tab:r1_stiff}}
\centering
\begin{tabular}{lll}
\toprule
\textbf{Quantity} & \textbf{Torque box} & \textbf{Skin}    \\ \midrule
$P$ [N]                                       & 1          & 1       \\
$L$ [m]                              & 1.325      & 1.385   \\ 
$\delta_{tip}$ [m]          & 4.7337e-3  & 0.51619 \\ 
$EI$ [Nm\textsuperscript{2}] & 163.8      & 1.72    \\ \bottomrule
\end{tabular}
\end{table}
Hence, the skin can be suitably discretised as lumped masses, modelling each strip as a separate mass, since it does not bring any appreciable stiffening contribution due to the use of the Agilus 30 strips between the Digital ABS sections. Thus, forty-seven lumped masses are added to the model as \verb|MASS21| elements with \verb|CERIG| links to the spar. 

\Cref{tab:7prop_s} {shows properties of the beam-like sections of interest, including moments of inertia }(I$_{xx}$ and I$_{yy}$) in m\textsuperscript{4}, areas (A) in m\textsuperscript{2} and torsion constants (J) in m\textsuperscript{4}. 

\begin{table}[!ht] 
\caption{Sections properties (Images not to scale).\label{tab:7prop_s}}
\centering
\begin{tabular}{llllll}
\toprule
\textbf{Section} & \includegraphics[width=2cm,keepaspectratio]{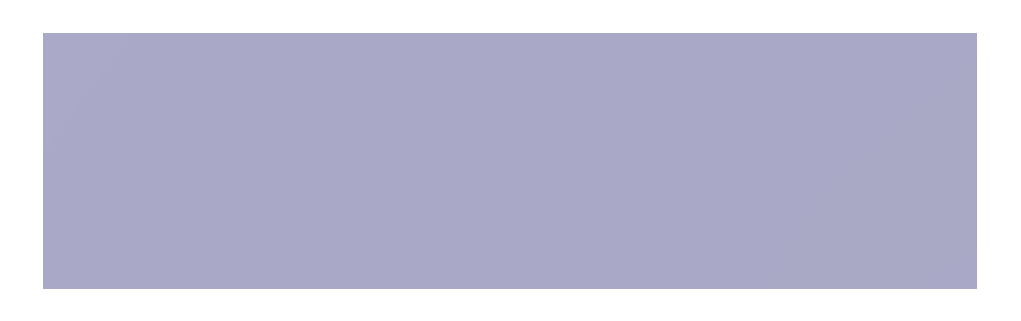}&\includegraphics[width=2cm,keepaspectratio]{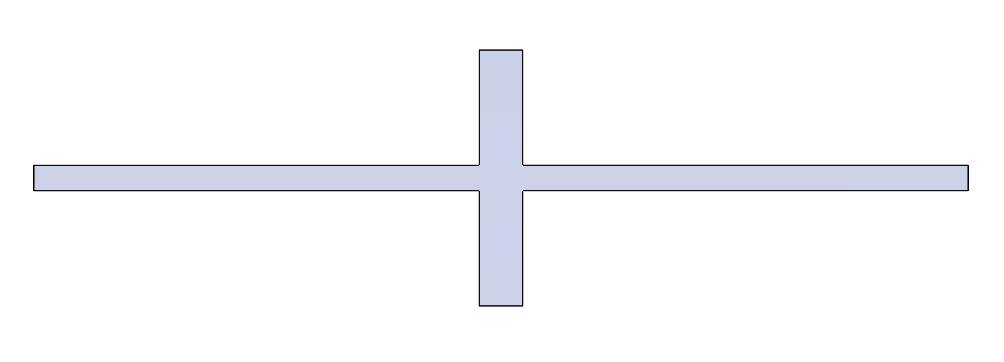} & \includegraphics[width=1.8cm,keepaspectratio]{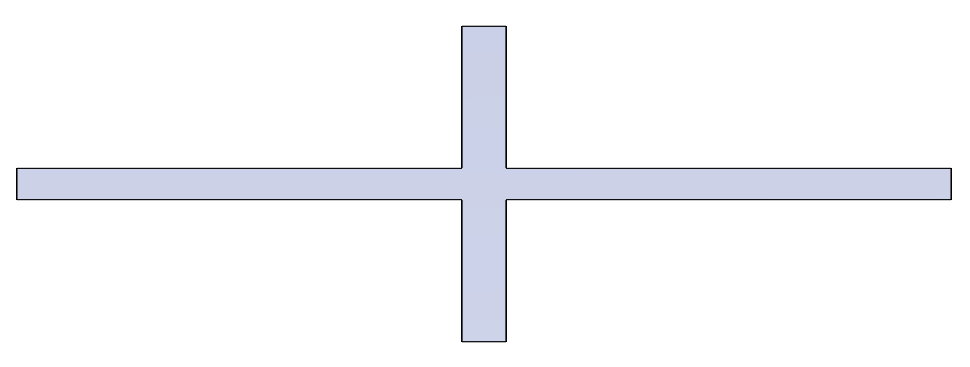} & \includegraphics[width=1.5cm,keepaspectratio]{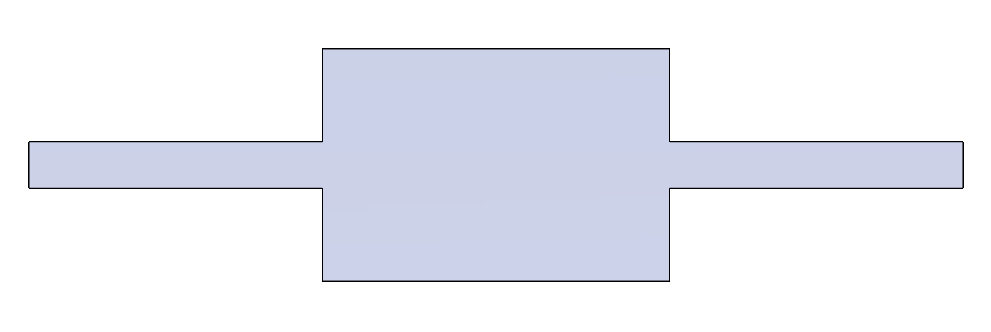} & \multicolumn{1}{c}{\includegraphics[width=.5cm,keepaspectratio]{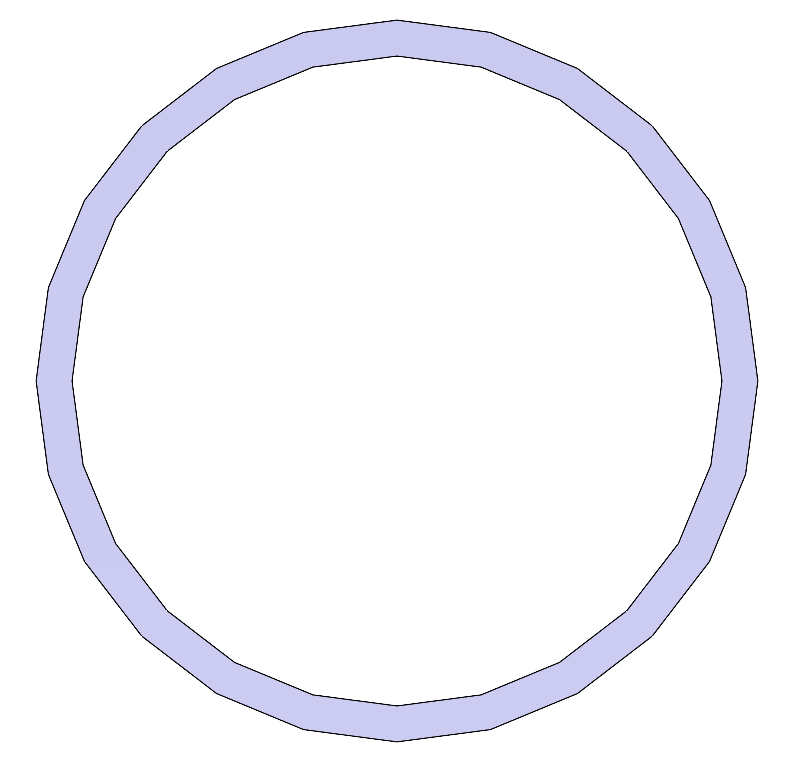}}\\\midrule
{I$_{xx}$} [m\textsuperscript{4}]	& 5$\times$10\textsuperscript{-8} & 6.50$\times$10\textsuperscript{-8} & 3.46$\times$10\textsuperscript{-8} & 1.29$\times$10\textsuperscript{-8} & 1.68$\times$10\textsuperscript{-10} \\
{I$_{yy}$} [m\textsuperscript{4}]	& 6.50$\times$10\textsuperscript{-7} & 2.30$\times$10\textsuperscript{-9} & 1.90$\times$10\textsuperscript{-9} & 1.24$\times$10\textsuperscript{-9} & 1.68$\times$10\textsuperscript{-10}\\
{A} [m\textsuperscript{2}]	& 14.60$\times$10\textsuperscript{-4}  & 2.07$\times$10\textsuperscript{-4} & 1.69$\times$10\textsuperscript{-4} & 1.98$\times$10\textsuperscript{-4} & 1.49$\times$10\textsuperscript{-11}\\
{J} [m\textsuperscript{4}]	& 7$\times$10\textsuperscript{-7}  & 4.5$\times$10\textsuperscript{-10} & 3.1$\times$10\textsuperscript{-10} & 3$\times$10\textsuperscript{-9} & 3.38$\times$10\textsuperscript{-10}\\
\bottomrule
\end{tabular}
\end{table}

{FEM validation and verification are performed for all developed models. Potential hidden constraints are assessed by inspecting residual forces and the strain energy associated with modal rigid-body motion. Connectivity between adjacent nodes is verified through a static analysis using a unit force load. Modal analysis of the free (unconstrained) model yields six rigid-body modes with frequencies below 0.005 Hz, and the ratio between the highest computed rigid-body frequency and the lowest elastic natural frequency is below 10\textsuperscript{-4}}\footnote{{The reference values are retrieved from a relevant standard for Space engineering. Specifically, the reader is referred to Structural {FEM}s ECSS-E-ST-32-03C standard}(\url{https://ecss.nl/standard/ecss-e-st-32-03c-structural-finite-element-models/}).}. {Concerning the mesh, a global element size of 0.001 m is adopted, since it is found not to influence the results (natural frequencies and stresses under a unit load applied at the tip) and does not adversely affect computational efficiency. The wing shear centre spanwise distribution is not considered as a possible model check, as the skin contribution to global rigidity is negligible compared to that of the torque box, such that the actual shear centre of the structure is defined solely by the latter, consisting of two parallel beams.}

{The FEM developed are then used to obtain modal parameters ($\omega_n$ and $\mathbf{\phi}_n$) by fully constraining the root of the specimens and running a damped modal analysis for extracting the modes between 0 and 150 Hz}. Rayleigh damping is used for the definition of the damping coefficients, which are derived from the experimental results in \cite{Dessena2022b}{, as shown in }\cref{sec:exp_fem}{. These models are used as a baseline for the FEMU procedure in later sections.}

\subsubsection{Experimental Setup\label{sec:7exp_s}}
In order to validate and, if needed, update the models, experimental data is needed. A previous testing campaign on the above-mentioned wings and sub-assemblies has been conducted in \cite{Dessena2022b} and {the} results are used in this work.

\begin{figure*}[!ht]
\centering
		{\includegraphics[width=.8\textwidth,keepaspectratio]{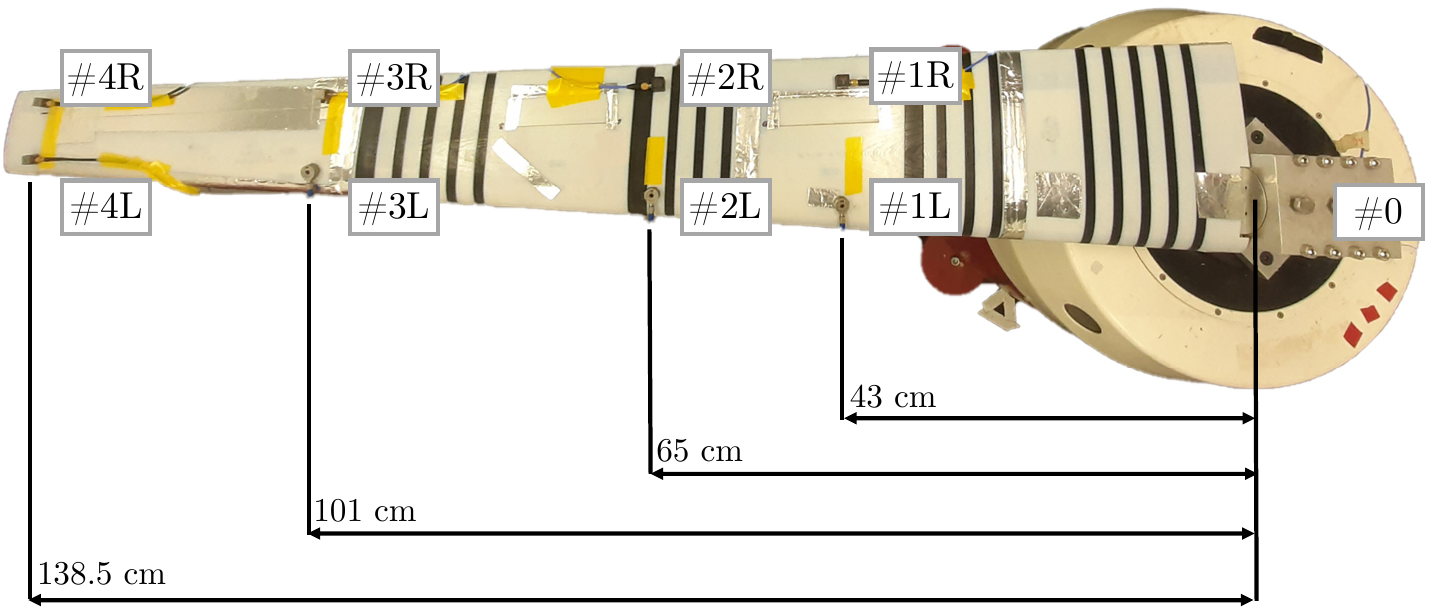}}
	\caption{{Accelerometer locations on the wing. The accelerometers may appear misaligned due to lens-induced perspective distortion in the camera image (adapted from }\cite{Dessena2022b})}
	\label{fig:fig2bis}
\end{figure*}

{A bandwidth-limited 2--400 Hz random verification at 0.305 g RMS (root mean square) input is carried out for the spar, torque box, and full wing assemblies using a shaker table, resulting in an 18-minute-long signal. Eight accelerometers, distributed on four rows across the span (channels \#1L--\#4R in }\cref{fig:fig2bis}{), are used to collect the vertical acceleration data only, due to equipment constraints. Hence, only vertical displacements and rotations can be inspected. The experimental setup consists of a Data Physics\textsuperscript{\textregistered} Signal Force\texttrademark{} modal shaker controlled through DP760\texttrademark{} closed-loop control software running on a consumer-grade laptop. An additional accelerometer (channel \#0 in }\cref{fig:fig2bis}{) is mounted on the clamp and used as feedback for the shaker control, while also serving as the reference channel for data processing. All accelerometers are connected to a National Instruments cDAQ-9178\footnote{\url{https://www.ni.com/es-es/shop/model/cdaq-9178.html?srsltid=AfmBOop6EI-DCHNnO_3Duot5TLTYNh-JkFfH5TP6tKjSi0a8gQumJqe5}}, and data are logged on a desktop workstation via an in-house \textsc{LabVIEW}\footnote{\url{https://www.ni.com/en-gb/shop/product/labview.html?srsltid=AfmBOorFBt8nWofzz9XhjNfwrytsD13QKLTRpevyQuOIK3FWFAQ3NvfK}} programme.  

The acquired time histories are sampled at $f_s=$ 512 Hz and band-pass filtered between 2.5 and 98 Hz to exclude low-frequency drift and high-frequency components outside the range of interest. Frequency response functions (FRFs) are computed by element-wise division of the FFT of each output channel by the FFT of the reference channel (channel \#0). A Savitzky--Golay filter is then applied to smooth the FRFs. The industry-standard Least Squares Complex Exponential (LSCE) method }\cite{Dezi2016,Spitznogle1970,Spitznogle1971} {is subsequently employed within a stabilisation diagram to identify the physical modes.}

The reader interested in the full experimental work, including details of the setup, acquisition system and post-processing, is referred to \cite{Dessena2022b,Dessena2022e}. 
{For further experimental work on the nonlinear dynamics of XB-2, readers are directed to }\cite{Dessena2022h}.

\subsubsection{Experimental Campaign and Preliminary Finite Element Models Results\label{sec:exp_fem}}
This subsection reports on the modal parameters extracted from the preliminary FEMs and compares them with those from the GVT campaign in \cite{Dessena2022b}.

In \cref{tab:7w_p}, the $\omega_n$ {obtained} {from the experimental campaign are compared with those found} from the damped modal analysis in Ansys Mechanical APDL 2021 R1, while only the $\zeta_n$ identified from the experimental campaign are shown. As usual, bending refers to modes with a predominant vertical (y-axis) displacement and lagging with a predominant horizontal (z-axis) displacement. The coupled modes considered in this study are coupled between bending and torsion (around the x-axis).

\begin{table*}[!ht] 
\caption{Natural frequencies identified from the experimental data and the preliminary FEMs.\label{tab:7w_p}}
\centering
\resizebox{\textwidth}{!}{
\begin{tabular}{lllllllll}
\hline
\multicolumn{9}{c}{\textbf{Natural Frequencies {[}Hz{]}}}                                                                                                                                                                                             \\ 
                                               \multicolumn{3}{c}{Spar} &                                                \multicolumn{3}{c}{Torque box}                                                & \multicolumn{3}{c}{XB-2 wing} \\\hline
\textbf{Mode}                 & Exp.      & FEM (\%)     & \textbf{Mode}                 & Exp.          & FEM (\%)       & \textbf{Mode}                 & Exp.         & FEM (\%)       \\ \hline
1\textsuperscript{st} Bending & 4.855     & 5.447        & 1\textsuperscript{st} Bending & 5.252         & 5.887          & 1\textsuperscript{st} Bending & 3.187        & 3.539          \\
                                               &           & (12.19)      &                                                &               & (12.10)        &                                                &              & (11.03)        \\
1\textsuperscript{st} Lagging & -         & 26.917       & 2\textsuperscript{nd} Bending & 25.933        & 30.617         & 1\textsuperscript{st} Coupled & 11.752       & -              \\
                                               &           &              &                                                &               & (18.07)        &                                               {(bending-torsion)} &              &                \\
2\textsuperscript{nd} Bending & 26.966    & 30.597       & 1\textsuperscript{st} Lagging & -             & 35.080         & 2\textsuperscript{nd} Coupled  & 17.447       & 17.774         \\
                                               &           & (13.46)      &                                                &               &                &        {(bending-torsion)}                                         &              & (1.88)         \\
1\textsuperscript{st} Torsion & -         & {68.049}   & 1\textsuperscript{st} Coupled & 76.242        & 85.162         & -                                              & -            & -              \\
                                               &           &              &    {(bending-torsion)}                                            &               & (11.70)        &                                                &              &                \\
3\textsuperscript{rd} Bending & 76.851    & 88.757       & -                                              & -             & -              & -                                              & -            & -              \\
                                               &           & (15.49)      &                                                &               &                &                                                &              &                \\ \hline
\multicolumn{9}{c}{\textbf{Damping Ratios {[}-{]}}}                                                                                                                                                                                                    \\
                                               \multicolumn{3}{c}{Spar} &                                                \multicolumn{3}{c}{Torque box}                                                & \multicolumn{3}{c}{XB-2 wing} \\\hline
\textbf{Mode}                 & Exp.      & FEM          & \textbf{Mode}                 & Exp.          & FEM            & \textbf{Mode}                 & Exp.         & FEM            \\ \hline
1\textsuperscript{st} Bending & 0.033     & -            & 1\textsuperscript{st} Bending & 0.022         & -              & 1\textsuperscript{st} Bending & 0.024        & -              \\
2\textsuperscript{nd} Bending & 0.010     & -            & 2\textsuperscript{nd} Bending & 0.014         & -              & 1\textsuperscript{st} Coupled & 0.047        & -              \\
3\textsuperscript{rd} Bending & 0.014     & -            & 1\textsuperscript{st} Coupled & 0.017         & -              & 2\textsuperscript{nd} Coupled & 0.038        & -              \\ \hline
\end{tabular}
}
\end{table*}

Notably, more modes are, generally, identified in the same frequency interval in the FEMs. \Cref{tab:7w_p} shows only the results for the modes identified between the first and last modes that can be found in both the experimental and FEMs data. At least three coincident modes are found for the spar {(the first three bending modes)} and torque box {(the first bending and the first two coupled modes)}; however, this is not the case for the full wing. Only two coherent modes are identified and, thus, they {are used} for the FEMU process. Hence, three modes are considered for the FEMU of the spar and torque box, while only two for the FEMU of the full wing. {It is worth noting that all coupled modes identified show couplig between bending and torsion.}

The $\zeta_n$ values shown in \cref{tab:7w_p} {are} found to obey classical Rayleigh damping. Hence, Rayleigh damping {is} assumed in the FEMs and the relative $\alpha$ and $\beta$ parameters {are} obtained by fitting the formulation below. 
\begin{equation}
    \bm{C} = \alpha\bm{M} + \beta\bm{K} \;\;\text{and}\;\; \zeta = \frac{1}{2}\left(\frac{\alpha}{\omega}+\beta\omega\right)
    \label{eq:7ray}
\end{equation}
\noindent where $\bm{C}$ {is the damping matrix}, $\alpha$ is the mass proportional Rayleigh damping coefficient, $\beta$ is the stiffness proportional Rayleigh damping coefficient, $\bm{M}$ {is the mass matrix}, $\zeta$ is the damping ratio and $\omega$ is the frequency in rads\textsuperscript{-1}. {$\alpha$ and $\beta$ are retrieved from the experimental results and used in the respective FEM. So damping coefficient or values are not used as updating variables.}
In terms of $\bm{\phi}_n$, the MAC values between the experimental and FEM results are above 0.9, {except} for the 1\textsuperscript{st} coupled mode of the torque box and the 2\textsuperscript{nd} coupled mode of the full wing. They stand at 0.8 and 0.75 respectively.  

Given the results outlined for the preliminary FEMs, the FEMU task is fundamental to guarantee a model that represents as closely as possible the real system. Particular attention should be paid to the parameters that are {furthest} from the experimental values, such as the above-mentioned $\bm{\phi}_n$ and most of the $\omega_n$.

\subsection{{Finite Element Model Updating of the Flexible Wing}}\label{sec:7femu}
The FEMU procedure follows a standard {approach} for surrogate-based iterative FEMU methods. In this section, the practical implementation of the point infill when max(EI) is reached in the MATLAB-based\footnote{{This was carried out using MATLAB 2021b (}\url{https://uk.mathworks.com/products/new_products/release2021b.html})} \texttt{rEGO} (as available in \cite{Dessena2022i}) and the Ansys Mechanical APDL model.

{The choice of using ANSYS Mechanical through APDL stems from the idea of having a flawless interface between the FEM software and MATLAB, where the rEGO-based optimisation routine runs. Additionally, APDL enable for the two-way (in and out) data flow between MATLAB and ANSYS Mechanical, such that the output parameters of interest can be retrieved by MATLAB, which in turn sends the input parameters for the model. This process is outlined below.}

\begin{figure*}[!ht]
\centering
	\begin{subfigure}[c]{.49\textwidth}
	\centering
		{\includegraphics[width=.9\textwidth,keepaspectratio]{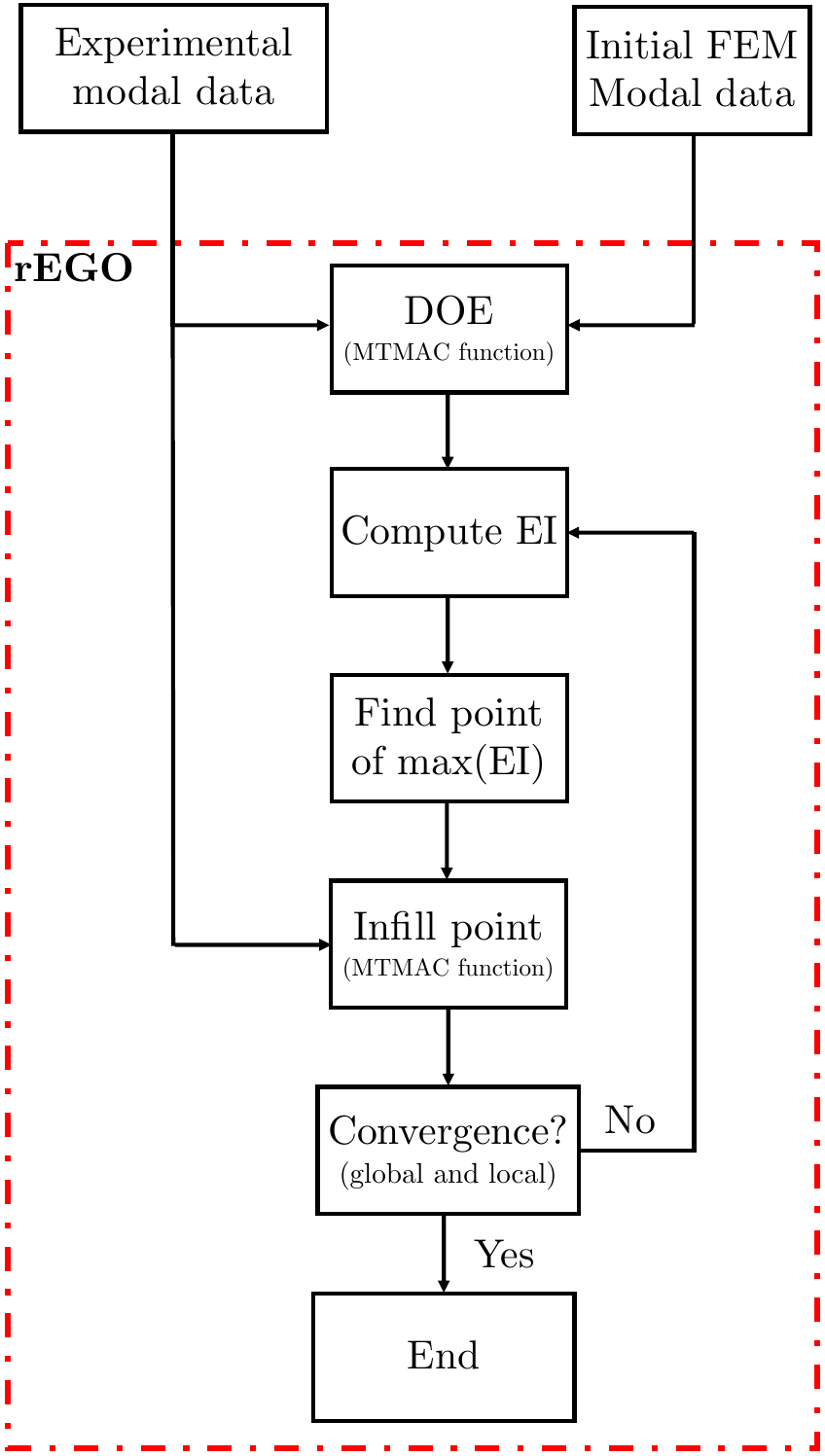}}
		\captionsetup{font={it},justification=centering}
		\subcaption{\label{fig:r1d1}}	
	\end{subfigure}
    \begin{subfigure}[c]{.49\textwidth}
	\centering
    \includegraphics[width=.725\textwidth,keepaspectratio]{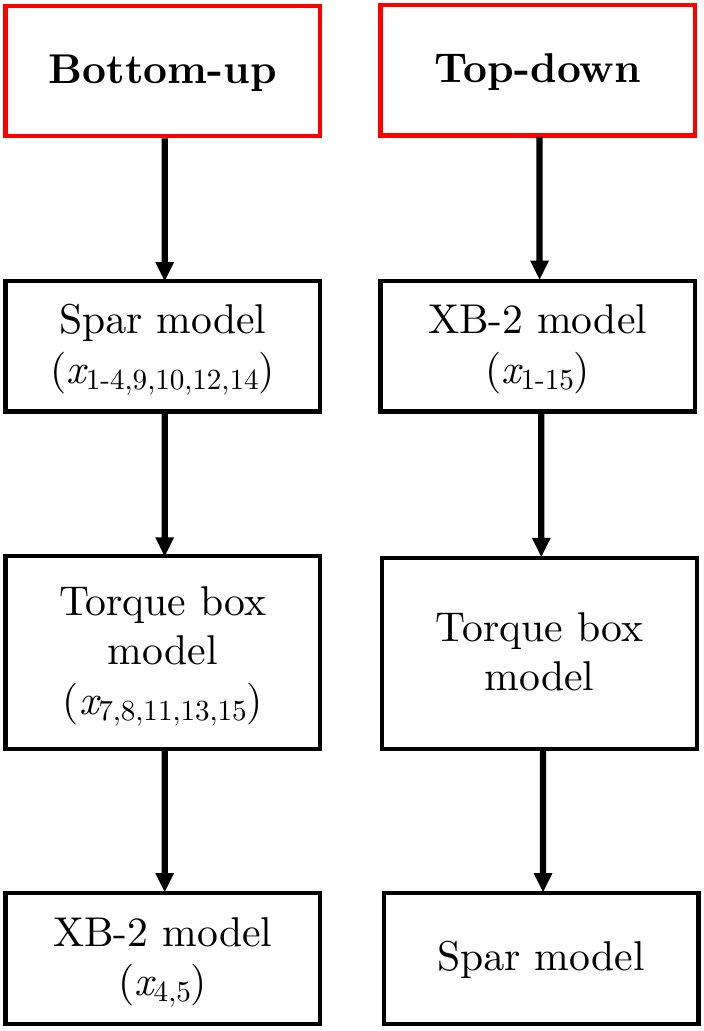}
		\captionsetup{font={it},justification=centering}
		\subcaption{\label{fig:r1d2}}	
	\end{subfigure}
	\caption{FEMU workflow: \cref{fig:r1d1} shows the optimisation workflow (in the red box the rEGO implementation) and \cref{fig:r1d2} the assembly-like structure and relative optimisation variables.}
	\label{fig:r1d}
\end{figure*}

First, a baseline discredited FEM is defined as an input file for Ansys Mechanical APDL. {The input file includes several parameters associated with the model properties. The MATLAB implementation retrieves the corresponding parameter set from rEGO and updates the input file accordingly. MATLAB then calls ANSYS Mechanical APDL to execute the modified input file. Once the run is completed, ANSYS generates an output file containing the FEM-based modal parameters. Then, this file is imported into MATLAB to compute the MTMAC and update the surrogate model.} This is shown in \cref{fig:r1d1}, while{,} for a practical implementation{, the interested reader can} refer to the \href{sec:6_data}{Data Availability Statement}, {where the full code implementation of this work is shared}. {However, it is important to note that for the scope of the validation of the proposed assembly-like FEMU method, any iterative optimiser of choice could replace rEGO in the red dashed-lined box in }\cref{fig:r1d1}.

A set of parameters needs to be defined to carry out the FEMU process. Modal properties are a direct function of material and geometric properties in {discretised} models. Hence, {their} tuning is considered in this work for the FEMU {task}. The density $\rho$ of all materials, as per \cref{tab:7prop}, is selected as an updating parameter. The same is done for the Young Modulus $E$ of all materials, apart from the Agilus 30 and the Digital ABS, because the skin is modelled as lumped masses (no stiffness). In addition, the geometric properties, such as $I_{yy}$ and J, are selected as optimisation variables. The lumped mass used to model the reinforcement plates at the spar mid-span is employed as updating parameters. {Lastly, a conceptual FEMU on the spar showed that the FEM would not converge adequately to the experimental results using only one material definition. Hence, the two halves of the spar are defined with two different material entries with base values for 6082-T6 aluminium.}

A total of fifteen parameters are identified across all the specimens for the FEMU process. For the {assembly-like} approach, the parameters are assigned to a given part or sub-assembly. For the spar FEMU, eight parameters are updated, while for the torque box thirteen and for the XB-2 model all fifteen are tuned. The parameters specific to the spar model include the two 6082-T6 Aluminium densities and $E$, the reinforcement plates lumped mass, and the $I_{yy}$ of the three spar sections ($x_{1-4,9,10,12,14}$). The torque box includes all those from the spar and adds, the $\rho$ and E for the stainless steel tube and the J of the three spar sections ($x_{7,8,11,13,15}$). Finally, the XB-2 wing inherits all the parameters from the torque box plus those relative to the modelling of the skin as lumped masses ($x_{4,5}$): the densities of Agilus 30 and Digital ABS. The updating parameters lower and {upper bounds ratios} are set at 0.6 and 1.4 and they actively scale the baseline value to match the FEM response to the experimental one. See \cref{tab:7param} for a breakdown of the variables to be updated and \cref{fig:r1d2} for a workflow of the model updating and data fetching.

\begin{table*}[!ht] 
\caption{{Properties updated for the FEMU processes.}\label{tab:7param}}
\centering
\resizebox{\textwidth}{!}{
\begin{tabular}{llll}
\toprule
\multicolumn{2}{c}{\textbf{Poperty}} & \textbf{Description} & \textbf{Model}\\\midrule
$x_1$ & 6082-T6 aluminium $\rho$&  $\rho$ of the inboard half of the spar. & Spar\\
$x_2$ & 6082-T6 aluminium E&  E of the inboard half of the spar. & Spar\\
$x_3$ & 6082-T6 aluminium $\rho$&  $\rho$ of the outboard half of the spar. & Spar\\
$x_4$ & 6082-T6 aluminium E&  E of the outboard half of the spar. & Spar\\
$x_5$ & Digital ABS $\rho$&  $\rho$ of the Digital ABS sections.& XB-2 wing\\
$x_6$ & Agilus 30 $\rho$&  $\rho$ of the Agilus 30 sections.& XB-2 wing\\
$x_7$ & Stainless steel $\rho$&  $\rho$ of the stainless steel tube.&Torque box\\
$x_8$ & Stainless steel E&  E of the stainless steel tube.&Torque box\\
$x_9$ & Reinforcement plates & Lumped mass value for the reinforcement plates discretisation.& Spar\\
$x_{10}$ & I$_{yy}$ spar (inboard) & Spar inboard cross-section I$_{yy}$ & Spar\\
$x_{11}$ & J spar (inboard) & Spar inboard cross-section J&Torque box\\
$x_{12}$ & I$_{yy}$ spar (inboard) & Spar mid-span cross-section I$_{yy}$& Spar\\
$x_{13}$ & J spar (inboard) & Spar mid-span cross-section J&Torque box\\
$x_{14}$ & I$_{yy}$ spar (inboard) & Spar outboard cross-section I$_{yy}$& Spar\\
$x_{15}$ & J spar (inboard) & Spar outboard cross-section J&Torque box\\\bottomrule
\end{tabular}}
\end{table*}

Since the main aim of this work is to investigate the goodness and feasibility of the {assembly-like} approach for the FEMU of complex structures, different updating scenarios have to be defined. Two approaches are determined. The first {considers} the spar FEM as the base for the {assembly-like} approach and the torque box and XB-2 wing models are built from there. This, known as the bottom-up approach, means that the spar model is first updated, and then FEMU is carried out on the torque box and the wing, by carrying over the parameters updated in the previous component. {On the other hand, the traditional approach} works the other way around, as a top-down approach. First, the XB-2 wing model is updated taking into consideration all fifteen parameters{. Then,} the torque box and spar model are built with the {fetched} parameters identified in that study. \Cref{tab:7ph} and \cref{fig:r1d2} recap the phases of the {assembly-like} approach.

\begin{table}[h] 
\caption{{Assembly-like} approach schematic.\label{tab:7ph}}
\centering
\begin{tabular}{llll}
\toprule
\multicolumn{3}{c}{\textbf{{Assembly-like} Approach}}\\ \midrule
           & Bottom-up (FEMU\_1) & Top-down (FEMU\_2)       \\
Spar & FEMU - 8 parameters & Fetching 8 parameters        \\
Torque box & FEMU - 5 parameters & Fetching 13 parameters \\
XB-2 wing  & FEMU - 2 parameters & FEMU - 15 parameters   \\ \bottomrule
\end{tabular}
\end{table}

{For completeness, the selected optimisation routine (shown in }\cref{fig:r1d1}{) , rEGO, is discussed herein. Originally, }rEGO \cite{Dessena2022c,Dessena2022d} is introduced to broaden the search capability of the Efficient Global Optimization (EGO) \cite{Jones1998,Forrester2008}. The main aim is to establish rEGO as a global-local technique, in a hybrid sense \cite{Keane2007}, able to both navigate search spaces globally (avoiding local minima) and land, not only, in the area of the global minimum but as close as possible to it, via the implementation of refinement and selection techniques. The main novelties of EGO are the use of a Kriging surrogate model and the implementation of a new infill metric, the expected improvement (EI). The EI can be defined as a measure of how much could the known minimum improve if a given point is added to the data pool. The EGO workflow is similar to other surrogate-based techniques: (i) the design space is searched strategically, usually with Latin hypercube sampling \cite{Forrester2007}, (ii) the absolute value of the EI is maximised to find the suitable infill point and (iii) the point is infilled. The process is iterated between (ii) and (iii) until convergence is reached. For {EGO}, usually, this happens when EI is less than 1\% {of} the objective function minimum \cite{Jones1998}. In order to improve this process, rEGO retains the same global structure but introduces two important principles: refinement and selection. 

In this work, rEGO is considered due to its superior computational performance when compared to GAs and EGO \cite{Dessena2022c,Dessena2022d}. {This means that a Kriging function, the same as per EGO, is used in this work as the surrogate model for the optimisation problem.}
A suitable {objective} function for FEMU via rEGO was identified in \cite{Dessena2022c} in the MTMAC, which compares the experimental and numerical modal data, $\omega_n$ and MAC values, thus $\bm{\phi}_n$. The MTMAC was proven in \cite{Dessena2022c} to outperform, computationally and precision-wise, other goal functions commonly used for FEMU.  Hence, 
the {opposite of the MTMAC product} is used in this work and its formulation is presented as:

\begin{equation}
 \text{MTMAC}=1-\prod^n_{i=1}\dfrac{\text{MAC}(\bm{\phi}_i^E,\bm{\phi}_i^N)}{\Bigg(1+\frac{|\omega_i^N-\omega_i^E|}{|\omega_i^N+\omega_i^E|}\Bigg)}   
 \label{eq:7MTMAC}
\end{equation}
where $n$ {denotes} the number of modes, superscript E$_exp$ for experimental data and superscript N$_num$ for numerical data.

{Intuitively, the product in }\cref{eq:7MTMAC}{ acts as a joint agreement term across modes: each modal contribution is close to unity only when both the corresponding MAC is high, and the frequency mismatch is small. As a result, a single poorly matched mode yields a small product and therefore a large MTMAC penalty, which promotes balanced improvements across all selected modes. Compared to additive objectives, this formulation reduces sensitivity to arbitrary weighting between frequency- and shape-based residuals, while improving accessibility beyond frequency-only objectives that do not constrain} $\mathbf{\phi}_n$. {This means that a value of 0 the MTMAC inverse shows perfect correlation.}

The reader interested in a more thorough review of EGO is referred to \cite{Jones1998,Forrester2007,Forrester2008} and to \cite{Dessena2022c,Dessena2022d} for the introductory work on rEGO. A MATLAB tutorial for rEGO can be found in \cite{Dessena2022i}.

 \section{Results}\label{sec:res}
\subsection{The Spar}
First, let us consider the results in terms of $\omega_n$ for the spar. In \cref{tab:7w_spar} 
the $\omega_n$ obtained from the experimental campaign (Exp.), preliminary FEM (FEM), FEM updated via bottom-up (FEMU\_1) and FEM updated via top-down (FEMU\_2) are compared.

\begin{table}[!ht] 
\caption{Natural frequencies identified from the experimental data, the preliminary FEM and the updated FEMs for the spar. MAC values (diagonal terms only) of the models {w.r.t.} the experimental data.\label{tab:7w_spar}}
\centering
\begin{tabular}{lllll}
\toprule
\multicolumn{5}{c}{\textbf{Natural Frequencies [Hz]}}\\
\textbf{Mode} & Exp. & FEM (\%)& FEMU - Bottom-up (\%)& FEMU - Top-down (\%)\\\midrule
1\textsuperscript{st} Bending  & 4.885 & 5.447 (12.19) & 4.855 (-) & 5.062 (4.26)\\
2\textsuperscript{nd} Bending  & 26.966 & 30.597 (13.46) & 26.864 (-0.38) & 31.775 (17.83)\\
3\textsuperscript{rd} Bending  & 76.851 & 88.757 (15.49) & 80.624 (4.91) & 95.269 (23.97)\\
\hline
\multicolumn{5}{c}{\textbf{MAC Values (of the diagonal) [-]}}\\
\textbf{Mode} & \multicolumn{2}{c}{FEM} & FEMU - Bottom-up (\%)& FEMU - Top-down (\%)\\\midrule
1\textsuperscript{st} Bending & \multicolumn{2}{c}{0.98} & 0.99 & 0.99\\
2\textsuperscript{nd} Bending & \multicolumn{2}{c}{0.94} & 0.98 & 0.97\\
3\textsuperscript{rd} Bending & \multicolumn{2}{c}{0.90} & 0.98 & 0.94 \\
\bottomrule
\end{tabular}
\end{table}

{The $\omega_n$ results for the bottom-up approach are consistent with the experimental identification.} The bottom-up approach provides the best performance, while the top-down approach proves to be the less accurate method in this instance. The top-down approach $\omega_n$ are higher than those identified by the preliminary FEM, apart from the first bending mode.

\begin{figure*}[!ht]
    \centering
    \includegraphics[width=\textwidth,keepaspectratio]{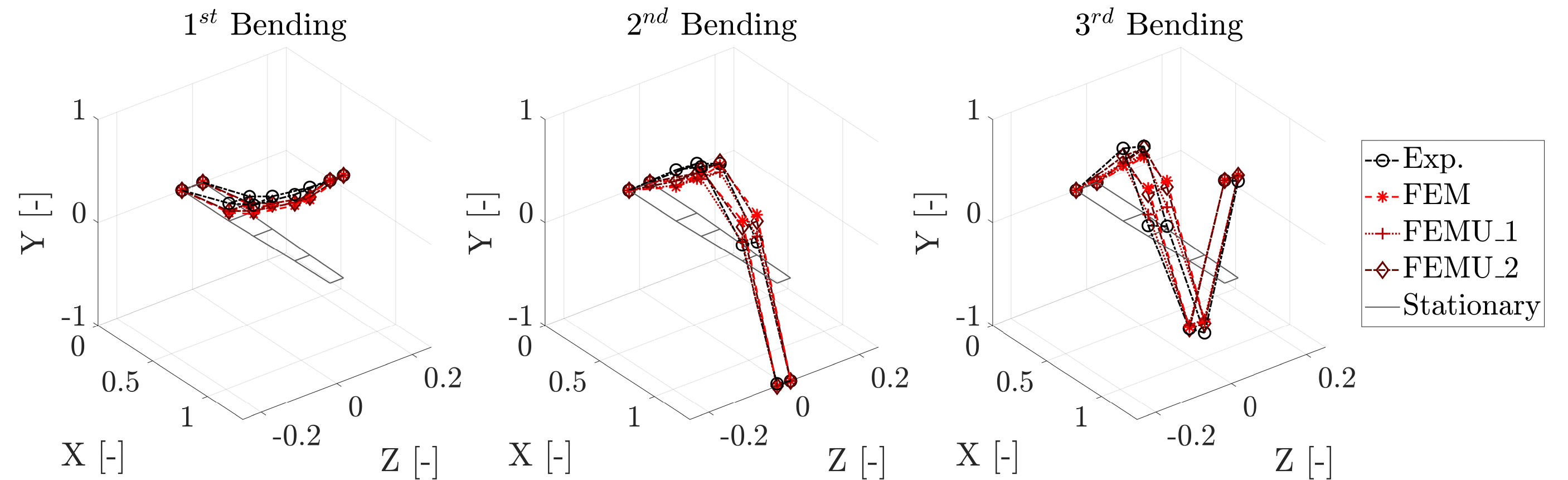}
    \caption{Mode shapes identified from the experimental data (Exp.), the preliminary FEM (FEM) and the updated FEMs (FEMU\_1-2) for the spar.\label{fig:7p_spar}}
\end{figure*}

The analysis on the identified $\bm{\phi}_n$ is {presented} in \cref{fig:7p_spar,tab:7w_spar}. The $\bm{\phi}_n$ from the experimental data, preliminary FEM and updated FEMs are superimposed for comparison in \cref{fig:7p_spar}, which is further investigated in \cref{tab:7w_spar}. There, the diagonal values of the MAC matrix are presented. Please note, the off-diagonal terms are not {shown} due to their negligible magnitude. 

Globally, all FEMs offer a good approximation of the spar $\bm{\phi}_n$ showing numerical correlation with values above 0.94 in \cref{tab:7w_spar} and graphical correlation in \cref{fig:7p_spar}. Nevertheless, the bottom-up approach performs better than the top-down approach and it is a clear improvement over the preliminary FEM. Notably, all the modes identified from the top-down approach are well correlated to the experimental case; however, they show less improvement than the bottom-up approach.

In order to compare the $\omega_n$ and $\bm{\phi}_n$ together, the penalty function of the optimisation routine can be considered. The MTMAC for the preliminary FEM is 0.30, for bottom-up is 0.07 and for top-down is 0.26. Clearly, bottom-up, globally, shows an improvement over the preliminary FEM. This means that the small deterioration in the $\omega_n$ is counterbalanced by a greater precision in the $\bm{\phi}_n$ identification. In terms of computational power, the top-down FEM takes the parameters from the FEMU of the XB-2 wing. However, bottom-up FEMU is carried out on the spar itself. The rEGO needs 327 model evaluations to converge to the presented results for the bottom-up approach.

\subsection{The Torque Box}
The $\omega_n$ identified for the torque box from the updated FEMs are compared to those from experimental data and the preliminary FEM in \cref{tab:7w_tor}. 

\begin{table}[!ht] 
\caption{Natural frequencies identified from the experimental data, the preliminary FEM and the updated FEMs for the torque box. MAC values (diagonal terms only) of the models {w.r.t.} the experimental data.\label{tab:7w_tor}}
\centering
\begin{tabular}{lllll}
\toprule
\multicolumn{5}{c}{\textbf{Natural Frequencies [Hz]}}\\
\textbf{Mode} & Exp. & FEM (\%)& FEMU - Bottom-up (\%)& FEMU - Top-down (\%)\\\midrule
1\textsuperscript{st} Bending  & 5.252 & 5.887 (2.76) & 5.658 (7.73) & 5.354 (1.94)\\
2\textsuperscript{nd} Bending  & 25.933 & 31.452 (21.18) & 32.663 (25.95) & 31.092 (19.89)\\
1\textsuperscript{st} Coupled  & 76.242 & 86.344 (13.25) & 82.992 (8.85) & 68.926 (-9.60)\\
\hline
\multicolumn{5}{c}{\textbf{MAC Values (of the diagonal) [-]}}\\
\textbf{Mode} & \multicolumn{2}{c}{FEM} & FEMU - Bottom-up & FEMU - Top-down \\\midrule
1\textsuperscript{st} Bending & \multicolumn{2}{c}{0.99} & 0.99 & 0.99\\
2\textsuperscript{nd} Bending & \multicolumn{2}{c}{0.96} & 0.95 & 0.96\\
1\textsuperscript{st} Coupled & \multicolumn{2}{c}{0.83} & 0.84 & 0.76 \\
\bottomrule
\end{tabular}
\end{table}

The $\omega_n$ identified from the preliminary FEM seems to be more closely related to those from experimental data. Their maximum error is 21.18\%, which is less than the maximum error for {the bottom-up FEM}. Notably, the maximum errors are always found for the second bending mode. On the other hand, the FEM from top-down has the lowest error (1.94\%) for $\omega_1$. This is a small improvement over the preliminary FEM, but it is much better than the bottom-up FEM (7.73\%). Until now, all the FEM approaches have overestimated the $\omega_n$ values. However, this changes with the first coupled mode identified by the top-down approach for the torque box, which underestimates the experimental value by 9.6\%. Concerning $\omega_3$, the bottom-up FEM shows an improvement over the preliminary FEM value, but the values for the other modes are overestimated. As it is clear from \cref{tab:7w_tor}, no model can be clearly set as the best, in terms of $\omega_n$ identification for the torque box.

\begin{figure*}[!ht]
    \centering
    \includegraphics[width=\textwidth,keepaspectratio]{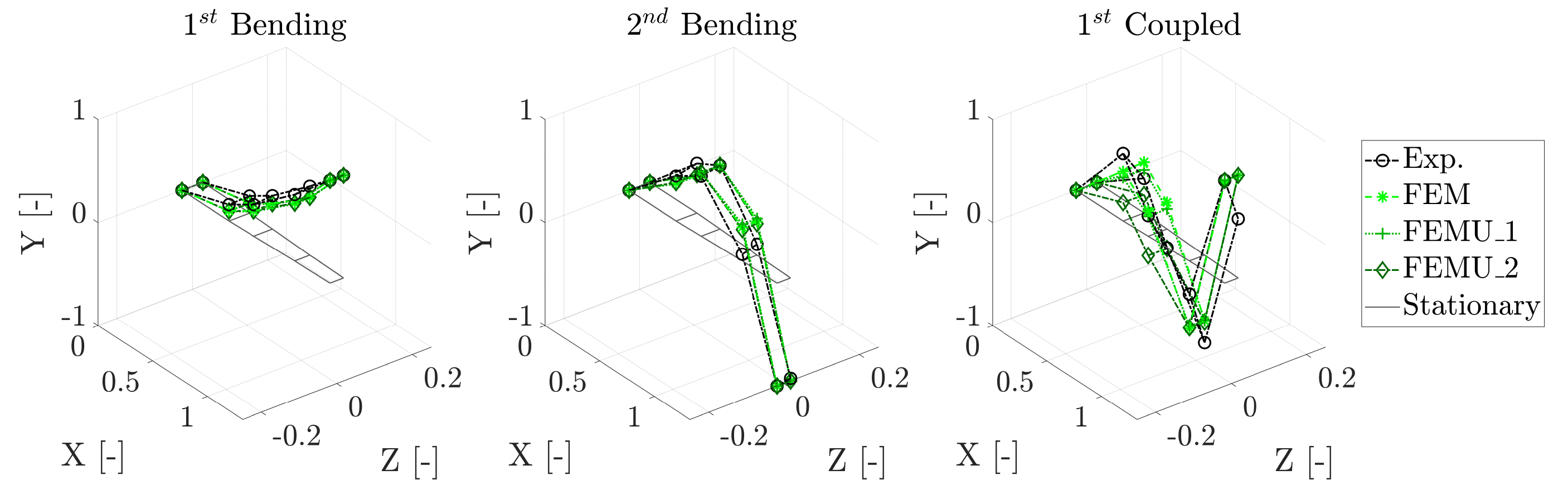}
    \caption{Mode shapes identified from the experimental data (Exp.), the preliminary FEM (FEM) and the updated FEMs (FEMU\_1-2) for the torque box.\label{fig:7p_tor}}
\end{figure*}
 
From \cref{fig:7p_tor}{, the most problematic $\bm{\phi}_n$ to reproduce is the first coupled mode. In particular, the considered FEMs underestimate the amplitude of the torsional rotation, indicating limited torsional fidelity of the adopted discretisation at the torque-box level. This is reflected in the MAC values, for which the lowest correlations are observed for the first coupled mode. The highest MAC for this mode is 0.84 for the bottom-up updated FEM, while the preliminary and bottom-up FEMs exhibit broadly similar trends across the remaining modes. The top-down updated FEM shows comparable behaviour for the lower modes, but the agreement deteriorates for the highest mode.

To assess $\omega_n$ and $\bm{\phi}_n$ simultaneously, the complement of MTMAC with respect to the experimental data is considered. The preliminary FEM yields ${\mathrm{MTMAC}}=0.34$, the bottom-up updated FEM ${\mathrm{MTMAC}}=0.35$, and the top-down updated FEM ${\mathrm{MTMAC}}=0.37$. The fact that the updated models do not outperform the preliminary FEM in this metric should be interpreted primarily as a \emph{model-form} limitation rather than a limitation of the updating strategy: The optimisation can only adjust parameters within the chosen discretisation, and a model that is mainly suited to out-of-plane dynamics may not capture twisting-dominated and coupled torsional behaviour with sufficient accuracy at the torque-box stage.

A more suitable representation of the torque-box torsional dynamics could be obtained through alternative discretisations, such as a shell-based torque-box model or a mixed beam--shell formulation (e.g. shells for webs and beams for stiffeners, intended as the long and short arms of the spar St. George's cross cross-section, respectively). These modelling choices are expected to improve the kinematic description of torsion and coupling, and would therefore provide a more informative basis for FEMU at this assembly level. Finally, in terms of computational effort, the bottom-up strategy required 780 model evaluations to reach convergence starting from the spar model.}

\subsection{The XB-2 wing}

\Cref{tab:7w_xb2} gives a quantitative overview of the precision, relative to the experimental data, of the parameters extracted via the FEMs. All FEM-extracted parameters are very close to the experimental data. The maximum error, relative to the experimental $\omega_n$, for the updated FEMs is 0.01\% in the bottom-up FEMU. 
\begin{table}[!ht] 
\caption{Natural frequencies identified from the experimental data, the preliminary FEMs and the updated FEM for the XB-2 wing. MAC values (diagonal terms only) of the models {w.r.t.} the experimental data.\label{tab:7w_xb2}}
\centering
\begin{tabular}{lllll}
\toprule
\multicolumn{5}{c}{\textbf{Natural Frequencies [Hz]}}\\
\textbf{Mode} & Exp. & FEM (\%)& FEMU - Bottom-up (\%)& FEMU - Top-down (\%)\\\midrule
1\textsuperscript{st} Bending  & 3.187 & 3.5397 (11.03) & 3.187 (-) & 3.187 (-)\\
2\textsuperscript{nd} Coupled  & 17.447 & 17.774 (1.88) & 17.448 (0.01) & 17.447 (-)\\
\hline
\multicolumn{5}{c}{\textbf{MAC Values (of the diagonal) [-]}}\\
\textbf{Mode} & \multicolumn{2}{c}{FEM} (\%)& FEMU - Bottom-up (\%)& FEMU - Top-down (\%)\\\midrule
1\textsuperscript{st} Bending & \multicolumn{2}{c}{0.98} & 0.99 & 0.99\\
2\textsuperscript{nd} Coupled & \multicolumn{2}{c}{0.75} & 0.77 & 0.78 \\
\bottomrule
\end{tabular}
\end{table}

\begin{figure*}[!ht]
    \centering
    \includegraphics[width=.8\textwidth,keepaspectratio]{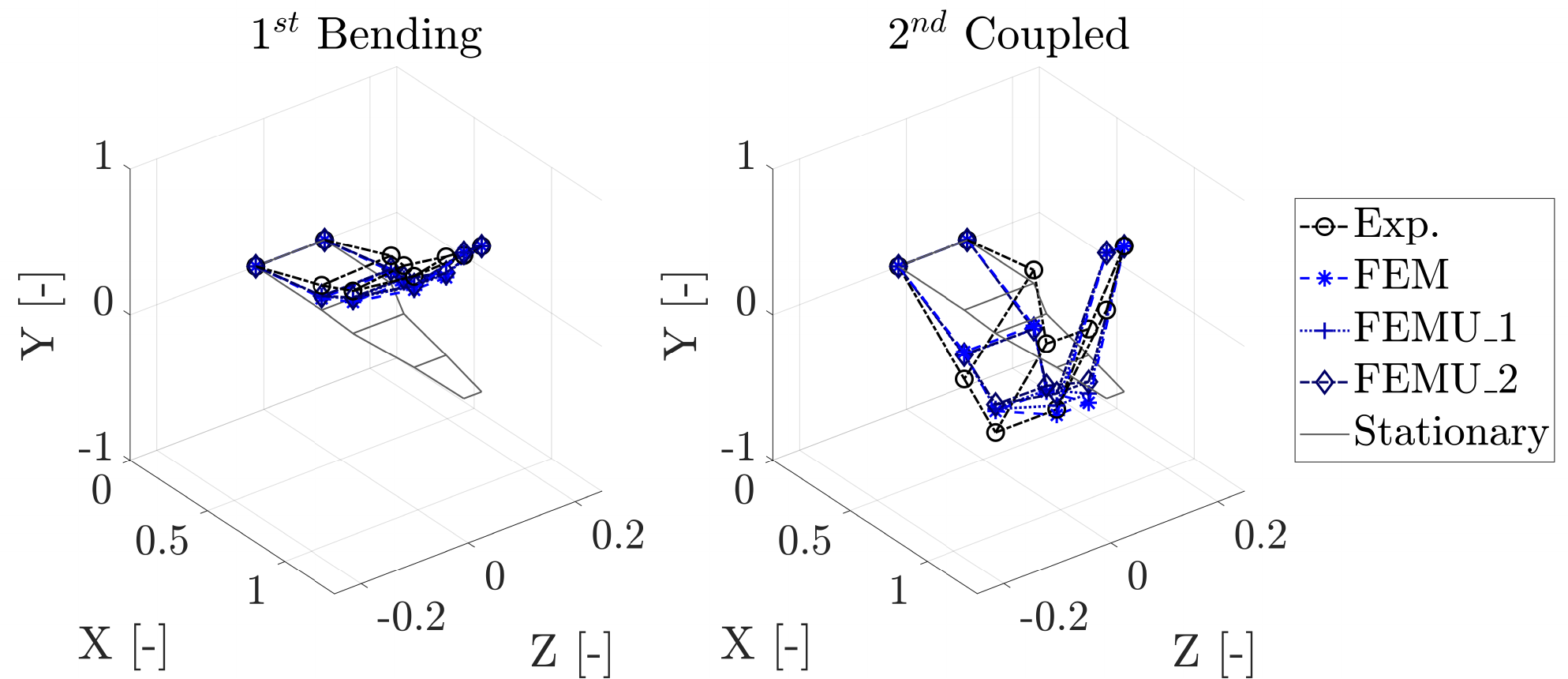}
    \caption{Mode shapes identified from the experimental data (Exp.), the preliminary FEM (FEM) and the updated FEMs (FEMU\_1-2) for the XB-2 wing.\label{fig:7p_xb2}}
\end{figure*}

In \cref{fig:7p_xb2} the first $\bm{\phi}_n$ is almost a perfect overlap for all FEMs. However, this does not happen for the second coupled mode, where, despite following the vertical deflection, the torsional magnitude is much smaller. These are confirmed in the quantitative analysis in \cref{tab:7w_xb2}, where the MAC values for the first bending mode are close to 1 for the updated FEMs and 0.98 for the preliminary model, but those for the second coupled mode do not go over 0.78, stopping at 0.75 for the preliminary FEM and 0.77 for the bottom-up FEM. This has to be redirected to the nature of discretisation, which is focused on the out-of-plane dynamics.

In order to give equal consideration to $\omega_n$ and $\bm{\phi}_n$ the MTMAC is addressed. The starting MTMAC for the XB-2 wing is 0.30, but the updated models are able to reduce to 0.24 (bottom-up) and 0.23 (top-down). In terms of percentages, the MTMAC of the bottom-up approach improved by 20\%, while the top-down approach improved by 23\%. Similar results in terms of precision are delivered by the bottom-up and 2 updated FEMs; however, the use of computational power needs to be taken into account. 

\section{Discussion}\label{sec:dis}

{In the following, computational cost is assessed using the number of FEM evaluations together with a proposed normalised, non-time-based index, the EFWE, which accounts for model size and solver memory footprint.}
Specifically,
\begin{equation}
\mathrm{EFWE}=\sum_{m} N_{\mathrm{eval},m}\left(\frac{N_{\mathrm{eq},m}}{N_{\mathrm{eq},\mathrm{XB2}}}\right)\left(\frac{M_{m}}{M_{\mathrm{XB2}}}\right),
\label{eq:efwe}
\end{equation}
{where $N_{\mathrm{eval},m}$ is the number of FE solver calls for model $m$, $N_{\mathrm{eq},m}$ is the solver-reported number of equations, and $M_m$ is the solver-reported total allocated memory.} \Cref{tab:efwe_models} {summarises these solver-reported proxies for the three FEM considered.}

\begin{table}[!ht]
\caption{{Model size and solver footprint used to compute EFWE.}\label{tab:efwe_models}}
\centering
\begin{tabular}{lccc}
\toprule
\textbf{Model} & \textbf{Nodes} & $N_{\mathrm{eq}}$ & $M$ [MB] \\
\midrule
Spar & 1451 & 7950  & 746.729 \\
Torque box & 2002 & 11238 & 768.238\\
XB-2 wing & 2049 & 11238 & 776.429\\
\bottomrule
\end{tabular}
\end{table}

{The top-down optimisation over 15 parameters took 1462 evaluations, all on the XB-2 wing model, and therefore yields $\mathrm{EFWE}=1462$. For the bottom-up approach, the evaluations at all preceding fidelity levels are accounted for: the spar model took 327 evaluations to converge, the torque box model converged after 780 evaluations and, lastly, the final update took 57 evaluations on the full wing, totalling 1164 evaluations. Therefore, the bottom-up strategy yields $\mathrm{EFWE}\approx 1051$, which is about 28\% lower than the top-down case.
Importantly, the most computationally expensive model (the full wing) is evaluated only 57 times in the bottom-up approach, while, by definition, it is evaluated 1462 times in the top-down approach (about $26\times$ fewer full-wing evaluations). This shows that, while the newly proposed bottom-up approach retains similar accuracy to the traditional approach, it also offers a lower computational burden when accounting for both evaluation counts and model complexity.}

\begin{figure*}[!ht]
    \centering
    \includegraphics[width=\textwidth,keepaspectratio]{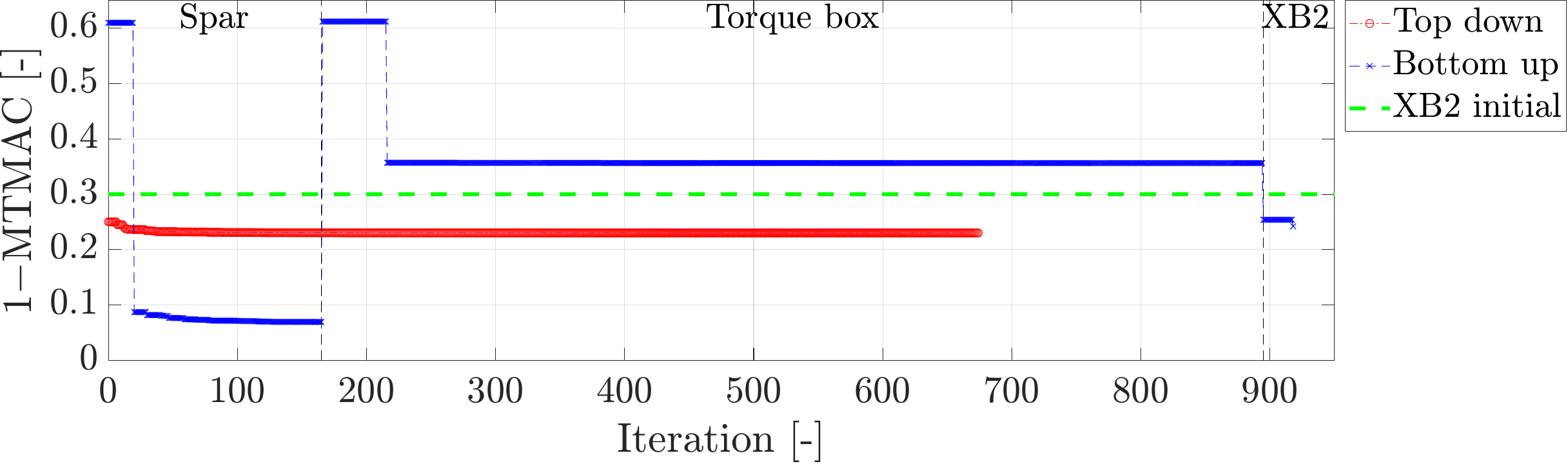}
    \caption{MTMAC complement value vs iteration.\label{fig:fig7}}
\end{figure*}

{Furthermore, on the performance benchmarking,} \cref{fig:fig7} {shows the MTMAC complement vs the iteration for the bottom-up and top-down cases. It is worth noting that the vertical dotted black lines and in-figure labels show the sub-assembly of reference of the bottom-up FEMU. Notably, more iterations are needed for the bottom-up approach. This might seem in opposition to what is stated about the number of model evaluations, which are lower for the bottom-up case. However, it should be recalled that at each iteration up to two model evaluations can be carried out when the $\epsilon_1$ is satisfied (see }\cref{fig:r1d}{). As expected, the MTMAC decreases as the number of iterations increases, and its final value is lower than the initial value for both FEMU strategies.}

In terms of model mass, the assembly-like method, as shown in \cref{tab:r1_mass}, improves the mass matching of the initial FEM, even if this was not the main objective of the optimisation (although $\rho$ is considered as a variable). On the other hand, the top-down FEM outperforms the two, at the cost of computational resources. However, this is a superficial analysis. It might seem that the top-down FEM is much better than the bottom-up FEM. However, this is not the case.  In fact, when the actual and model masses for the torque box and skin are compared separately, the absolute errors (on the components) are similar for the two methods, with the newly proposed approach showing slightly better performance for the skin mass (6.80 vs 6.98\%). {These small (less than {8.5}\%) differences are to be attributed to to the massless discretisation of the spar-to-tube links and to the uncertainty of the 3D-printed plastics density.} {Furthermore, they are sufficiently small and distributed (the errors absolute values for the torque box and skin in the two approaches are similar), which do not constitute a problem for potential further aeroelastic analyses and are a clear improvement {w.r.t.} the initial model.} Nevertheless, it is important to consider that the models developed here are {discretised} FEMs, such that the priority is to match the real system dynamics rather than the exact stiffness and mass values \cite{Cecrdle2022}.

\begin{table}[!ht] 
\caption{Identified models mass.\label{tab:r1_mass}}
\centering
\begin{tabular}{llll}
\toprule
 & \multicolumn{3}{c}{\textbf{Mass} [kg] (relative error {w.r.t.} experimental - \%)}   \\ 
\textbf{XB-2 Model} & {Torque box} & {Skin} & {Total} \\\midrule
Experimental & 1.362 & 1.662 & 3.024 \\ 
Initial FEM&1.224 (-10.13)&1.458 (-12.27) & 2.682 (-11.31) \\ 
Top-down FEM&1.250 (-8.22)&1.778 (+6.98)&3.029 (+0.02)\\ 
Bottom-up FEM&1.250 (-8.22)&1.549 (-6.80)&2.799 (-7.44)\\ \bottomrule
\end{tabular}
\end{table}

{These results highlight the computational advantage of the proposed assembly-like method and support the authors’ claims that a paradigm shift towards assembly-like methods should be considered in FEMU. Although the present study uses GVT-style modal information and MTMAC as the optimisation objective, the approach is not restricted to GVT and can be applied whenever suitable experimental data are available for model updating (e.g., ambient vibration testing, taxi vibration testing, flight vibration testing, and propeller-driven vibration testing). Moreover, by adopting an alternative optimisation objective, the same workflow can be extended to time-domain or frequency-domain data; however, these can be more sensitive to measurement noise and operational variability and may require additional preprocessing and uncertainty handling. Nevertheless, such approaches have been shown to be applicable in the literature} \cite{Yang2017}{.
On the other hand, this approach might not be suitable for non-assemblable structures, such as buildings (although some new construction techniques could be considered to be assemblies and should be investigated). Nevertheless, other applicable examples suitable for assembly-like approaches could be small satellites, such as CubeSats, and other mechanical systems, such as aero-engines. Complications might arise for systems that require altering the experimental setup to test sub-components. Nevertheless, this approach is scalable to larger assemblies, provided that the structure and subsystems remain testable. A potential use case in aerospace could be that of smaller launch vehicles, for which it is difficult to obtain GVT data due to their size, but subassemblies are testable. Furthermore, this study has only included the modelling of the system linear dynamics, thus any nonlinearity in the system is not considered in the FEM or FEMU process. Finally, a change of the modelling strategy and optimisation objective could potentially take these into account, but further validation would be needed, as the superposition principle is not directly applicable in the nonlinear regime.}

\section{Conclusions\label{sec:7con}}
In this work, a flexible wing finite element model {(FEM) }is characterised and updated using modal data from an experimental campaign following two assembly-like, bottom-up and top-down, approaches. {The main findings of this study can be condensed into the following points:}

\begin{itemize}
    \item A framework for assembly-like {FEM} updating is proposed and applied to a flexible wing;
    \item The direct approach (top-down) and the assembly-like approach (bottom-up) have similar precision in modelling the flexible wing (3\% difference in MTMAC and similar model mass estimation precision);
    \item {The bottom-up approach is more computationally efficient: when the evaluation counts are weighted by solver-reported model size and memory footprint, the resulting normalised full-wing-equivalent workload is about 28\% lower than for the top-down strategy;}
    \item {The top-down approach performs all evaluations on the most expensive full-wing model (1462 full-wing solves), whereas the bottom-up approach limits full-wing solves to 57 out of 1164 total evaluations (about 4.9\%), i.e., roughly 26 $\times$ fewer full-wing evaluations.}
 \item To the authors' knowledge, this is the first time that refined Efficient Global Optimisation is applied to a complex structure {FEM} updating;
    \item It is the first updated {FEM} available in the literature for the benchmark wing.
\end{itemize} 
Thus, the findings of this work support the paradigm shift of {FEM} updating towards assembly-like approaches, which is the recommended approach for similar structures{, although its main limitation is the need for sub-assembly modal data.}

\backmatter








\section*{Declarations}
\begin{itemize}
\item Funding: The authors from Cranfield University disclosed receipt of the following financial support for the research, authorship, and/or publication of this article: This work was supported by the Engineering and Physical Sciences Research Council (EPSRC) [grant number 2277626].
\item Conflict of interest/Competing interests: Gabriele Dessena is serving as an Editorial Board Member for the journal and has not been involved in the peer-review process for this article in any way.
\item Ethics approval and consent to participate: Not applicable.
\item Consent for publication: All authors have approved the final manuscript and agree to its submission.
\item Data availability: Data supporting (ANSYS Mechanical APDL input files and MATLAB scripts) this study are openly available from the Zenodo repository at [\url{https://doi.org/10.5281/zenodo.14025503}] under the terms of [GNU General Public License (GPLv3)].
\item Materials availability: Not applicable.
\item Code availability: Data supporting (ANSYS Mechanical APDL input files and MATLAB scripts) this study are openly available from the Zenodo repository at [\url{https://doi.org/10.5281/zenodo.14025503}] under the terms of [GNU General Public License (GPLv3)].
\item Author contribution: Conceptualisation, G.D.; methodology, G.D.; software, G.D.; validation, G.D.; formal analysis, G.D.; investigation, G.D.; resources, D.I., J.W., A.P., and L.Z.; data curation, G.D.; writing---original draft preparation, G.D.; writing---review and editing, G.D., D.I., J.W. A.P., and L.Z.; visualisation, G.D.; supervision, D.I., J.W., A.P., and L.Z.; funding acquisition, L.Z..
\end{itemize}

\end{document}